\documentclass[usenatbib,psfig]{mnras}

\usepackage[T1]{fontenc}

\DeclareRobustCommand{\VAN}[3]{#2}
\let\VANthebibliography\thebibliography
\def\thebibliography{\DeclareRobustCommand{\VAN}[3]{##3}\VANthebibliography}
\usepackage{amsmath,amsfonts,amssymb}
\usepackage{color}

\usepackage{times}
\usepackage{url}
\usepackage{float}

\usepackage{epsfig}
\usepackage{graphicx}
\usepackage{subcaption}
\usepackage[export]{adjustbox}

\usepackage{caption}
\usepackage{changepage}
\usepackage[justification=centering]{caption}
\usepackage{natbib}
\usepackage{wrapfig}

\usepackage{dirtytalk}
\usepackage{amssymb}
\usepackage{placeins}
\usepackage{threeparttable}
\usepackage{color}
\usepackage{float}
\newcommand{\cha}[1]{\textcolor{black}{#1}}

\newcommand{\referee}[1]{\textcolor{black}{#1}}

\usepackage[normalem]{ulem}
\usepackage{dblfloatfix}
\usepackage{multicol}
\usepackage[version=4]{mhchem}
\usepackage{titlesec}

\newcommand{\Msun}{\ensuremath{\textrm{M}_{\odot}}}

\newcommand{\Lsun}{\ensuremath{\textrm{L}_{\odot}}}

\newcommand{\Kms}{km~s$^{-1}$} 
\newcommand{\kms}{km\hspace{0.25em}s$^{-1}$}

\newcommand{\OI}{\mbox{O\hspace{0.25em}{\sc i}}}

\newcommand{\CII}{\mbox{C\hspace{0.25em}{\sc ii}}}

\newcommand{\MgII}{\mbox{Mg\hspace{0.25em}{\sc ii}}}

\newcommand{\SII}{\mbox{S\hspace{0.25em}{\sc ii}}}

\newcommand{\SiII}{\mbox{Si\hspace{0.25em}{\sc ii}}}
\newcommand{\SiIII}{\mbox{Si\hspace{0.25em}{\sc iii}}}

\newcommand{\CaII}{\mbox{Ca\hspace{0.25em}{\sc ii}}}

\newcommand{\TiII}{\mbox{Ti\hspace{0.25em}{\sc ii}}}

\newcommand{\FeII}{\mbox{Fe\hspace{0.25em}{\sc ii}}}
\newcommand{\FeIII}{\mbox{Fe\hspace{0.25em}{\sc iii}}}
\newcommand{\CoII}{\mbox{Co\hspace{0.25em}{\sc ii}}}
\newcommand{\CoIII}{\mbox{Co\hspace{0.25em}{\sc iii}}}

\newcommand{\Fefs}{$^{56}$Fe}
\newcommand{\Feff}{$^{54}$Fe}
\newcommand{\Cofs}{$^{56}$Co}
\newcommand{\Nifs}{$^{56}$Ni}
\newcommand{\Nife}{$^{58}$Ni}

\newcommand{\Dm}{\ensuremath{\Delta m_{15}(B)}}

\newcommand{\lam}{$\lambda$}
\newcommand{\llamb}{$\lambda$$\lambda$}

\newcommand{\chris}[1]{\textcolor[rgb]{0, 0, 0}{#1}}
\newcommand{\stefan}[1]{\textcolor[rgb]{0, 0, 0}{#1}}
\newcommand{\takashi}[1]{\textcolor[rgb]{0, 0, 0}{#1}}

\newcommand{\eg}{e.g.,\ }
\newcommand{\ie}{i.e.,\ }

\def\gsim{\mathrel{\rlap{\lower 4pt \hbox{\hskip 1pt $\sim$}}\raise 1pt \hbox {$>$}}}
\def\lsim{\mathrel{\rlap{\lower 4pt \hbox{\hskip 1pt $\sim$}}\raise 1pt \hbox {$<$}}}
\def\gtaprx {\lower .1ex\hbox{\rlap{\raise .6ex\hbox{\hskip .3ex
	{\ifmmode{\scriptscriptstyle >}\else
		{$\scriptscriptstyle >$}\fi}}}
	\kern -.4ex{\ifmmode{\scriptscriptstyle \sim}\else
		{$\scriptscriptstyle\sim$}\fi}}}
\def\ltaprx {\lower .1ex\hbox{\rlap{\raise .6ex\hbox{\hskip .3ex
	{\ifmmode{\scriptscriptstyle <}\else
		{$\scriptscriptstyle <$}\fi}}}
	\kern -.4ex{\ifmmode{\scriptscriptstyle \sim}\else
		{$\scriptscriptstyle\sim$}\fi}}}

\usepackage{graphicx}	
\usepackage{amsmath}	

\title[SN\,2012dn]{Abundance Stratification in Type Ia Supernovae -- VIII. The 03fg-like SN\,2012dn: a double-degenerate merger interpretation}

\author[Aouad et al]{Charles J. Aouad$^1$\thanks{E-mail:C.J.Aouad@2020.ljmu.ac.uk}, 
Paolo A. Mazzali$^{1,2}$,
Stephan Hachinger$^{3}$,
Chris Ashall$^{4}$
\\
\\
$^1$ Astrophysics Research Institute, Liverpool John Moores University, 146 Brownlow Hill, Liverpool L3 5RF, UK\\
$^2$ Max-Planck Institut f\"{u}r Astrophysik, Karl-Schwarzschild-Str. 1, D-85748 Garching, Germany\\
$^3$ Leibniz Supercomputing Centre (LRZ) of the BAdW, Boltzmannstr. 1, D-85748 Garching, Germany\\
$4$ Institute for Astronomy, University of Hawai'i at Manoa, 
2680 Woodlawn Dr., Hawai'i, HI 96822, USA\\
}

\date{Accepted 2026 September 4. Received 2026 July 17; in original form 2026 April 27}

\pubyear{2023}

\begin{document}
\label{firstpage}
\pagerange{\pageref{firstpage}--\pageref{lastpage}}
\maketitle
\begin{abstract}
   A detailed study using abundance tomography of SN\,2012dn, a peculiar Type Ia supernova, is presented. Despite exhibiting a normal peak luminosity, it retains early-phase super-Chandrasekhar (03fg-like) characteristics: weak \FeIII, narrow intermediate-mass element (IME), persistent carbon, and the absence of high-velocity \ion{Ca}{ii} features. Its nebular spectrum is unusually faint due to grey dimming beginning $\sim$60 days after maximum and shows [\OI] emission, rare in SNe\,Ia. While a Chandrasekhar-mass density profile reproduces the photospheric phase, it fails at late times. The [\OI] emission requires additional low-velocity mass, implying a total ejecta mass of $1.66$ \Msun, including $0.33$ \Msun\ of oxygen (with $0.1$ \Msun\ in the core). The \Nifs\ mass is estimated to be $0.45$--$0.49$ \Msun, a range set by uncertainties in the grey-extinction correction applied at the nebular phase, and is insufficient on its own to account for the peak luminosity. Stable iron is confined to intermediate layers and absent from both the core and outer ejecta, the latter indicating sub-solar progenitor metallicity. Silicon and sulfur span the full ejecta, while carbon extends down to $v \sim 6000$\kms. These properties favour a double-degenerate CO–CO white dwarf merger scenario. The remaining luminosity deficit may be explained by additional energy input from a weak interaction with a low-mass, carbon-rich circumstellar shell.
\end{abstract}


\section{INTRODUCTION}

Supernovae Type Ia (SNe Ia) are widely believed to result from the thermonuclear disruption of carbon-oxygen white dwarfs in  binary systems \citep{hillebrandt2000, mazzali2007}, and their width-luminosity relation has made them essential tools in cosmology \citep{Phillips1993, Perlmutter1998, Riess1998}. However, several aspects of these phenomena remain under debate. The exact nature of the progenitor system is still uncertain, with studies unable to conclusively determine whether single or multiple evolutionary channels are responsible \citep{whelan_SD_scenario, Iben_1984_DD_scenario, livio_2018_progenitors_of_Ia}. Similarly, it remains unresolved whether the observed diversity stems from different white dwarf masses, varying explosion mechanisms, or a combination of both. Furthermore, key physical processes, such as the exact conditions for carbon ignition and the specific mechanisms of flame propagation, remain poorly constrained \citep{hillebrandt2000, hillebrandt_2013_reviewarticle}. There is also a significant variation in peak luminosity among different SNe\,Ia (spanning $\sim$2--3 magnitudes in the $B$ band; e.g., \citealt{timmes2003_metallicity, Li_2011_lum_functionof_SNIa, ashall_2016_Lum_dist_of_Ia, Ashall2016}). This dispersion is attributed to the varying amounts of \Nifs\ produced during the explosion, but the reasons for this variation are not yet clear \citep{mazzali2001}. Furthermore, the wide range of spectroscopic features and their diversity remain difficult to understand despite numerous classification attempts \citep{nugent1995_Rsi, benetti2005, Branch_classification_2006, wang_classifc_2009}.

The event begins with explosive carbon burning when a threshold density of $\sim$ 2$\times$10$^9$g\,cm$^{-2}$\,is attained \citep{iwamoto1999}, triggering nuclear fusion that starts with the synthesis of the heaviest elements at the core. As the burning front moves outward into lower-density regions, fusion produces progressively lighter elements. This results in a layered ejecta structure \citep{mazzali2007}: a core of neutron-rich \Nife\ and \Feff\ is surrounded by radioactive \Nifs, followed by a layer of intermediate-mass elements (IMEs) like silicon and sulfur. The outermost layers consist of oxygen and carbon \citep{Parrentcarbonfeatures2011, silverman_carbon-features}, representing incomplete burning or original white dwarf material.

The freshly synthesized \Nifs\ decays through the  \Nifs~$\rightarrow$~\Cofs~$\rightarrow$~\Fefs\  chain \citep{pankey1962PhDT, Colgate_Mckee_1969_nifsdecay} releasing $\gamma$-rays and positrons that thermalise  with the expanding ejecta, producing optical photons. These photons escape when opacity decreases as the ejecta dilute, constituting the main power source of their light curve. 

The majority of events are classified as normal, characterized by relatively homogeneous observables \citep{filippenko_1997_optical_spectra_of_Sn}, typically exhibiting absolute magnitudes in the $B$-band of ($\mathrm{M}_B$) $-19.1 \pm 0.06$ mag \citep{Ashall2016}. Their early spectra display multiple lines of singly ionized IMEs, mainly \SiII\ \lam\ 6355, \SII, \MgII, \CaII, and neutral O. As the spectra evolve, Fe-group elements  are revealed and ions such \FeII\ and \FeIII\ start appearing. These events \stefan{largely overlap with the spectroscopic} ``core normal'' \stefan{or ``broad-line'' SN\,Ia classes} according to the classification scheme of \citet{Branch_classification_2006}, \stefan{and fall within the low- or high-velocity-gradient (LVG/HVG) groups} in the scheme of \citet{benetti2005}.

Sub-luminous events, termed 91bg-like, are 1.5–2.5 mag fainter than normal SNe\,Ia, reaching peak $M_B$ values as low as $\sim -16.85$ \citep[\eg\ SN\,1991bg;][]{taubenberger_underluminous_05bl_91bg, Ashall2016}. These underluminous transients are characterized by strong [\OI] and \TiII\ absorption lines near 7400 and 4450 \AA, respectively. They are classified as ``cool'' by \citet{Branch_classification_2006} or ``faint'' in the scheme of \citet{benetti2005}.

Hot, bright events, termed 91T-like \citep{obrien2023_1991TlikeSNe, phillips2024_1991T_like_SNe}, are typically about 0.5 mag brighter than normal events \citep{avichai_2017_handbookofsn_obser_class_of_SN} and are characterized by weak early \SiII \lam\ 6355, strong \FeIII\ lines and weak \CaII\ H$\&$K\ \citep{mazzali1995, Aouad_99aa, Aouad_iPTF16abc}. These events belong to the \stefan{``shallow silicon''} group according to \citet{Branch_classification_2006}, or \stefan{are part of the LVG group} according to \citet{benetti2005}.

\begin{figure}
\includegraphics[trim={15 4 40 20},clip,width=0.45\textwidth]{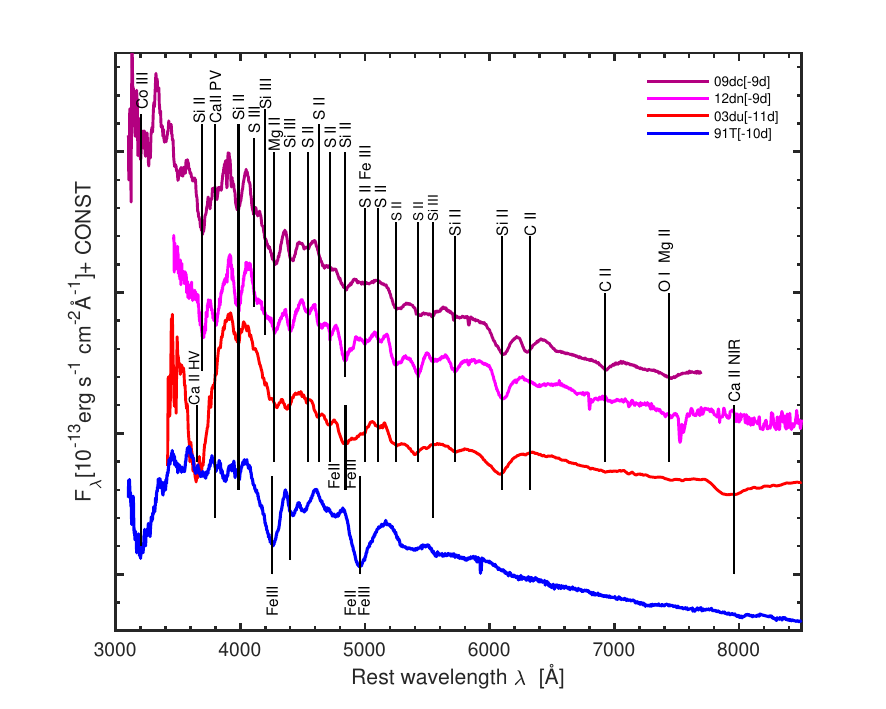}
 \caption{The early spectra of the 03fg-like SNe 2012dn and 2009dc compared to SN\,1991T, and the normal SN\,2003du,  \chris{highlighting their distinct IME-dominated features and weak \FeIII\ signatures discussed in the text.}}  
\label{spect_comp_early}
\end{figure}

\begin{figure}
\includegraphics[trim={15 4 40 20},clip,width=0.45\textwidth]{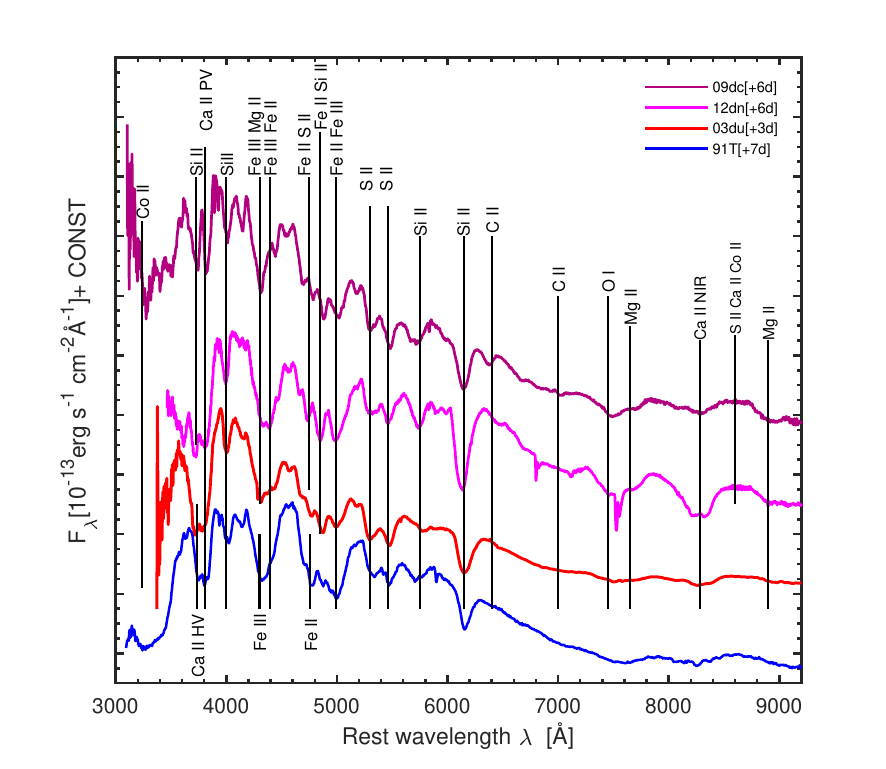}
 \caption{Spectra one week after $B$ maximum of the 03fg-like, SNe 2012dn, 2009dc compared to SN 1991T, and the normal SN\,2003du.}  
\label{spect_comp_late}
\end{figure}

Another peculiar set of events has emerged, \referee{classified as 2003fg-like \citep{SN2003fg, ashall_2021_2003fg_superchandra,  Lu_ASASSN-15hy, Dimitriadis_2022_2020esm_superchandra, Siebert_2023_2020hvf, nagao_polarization_03fg_like, Bose_2026_2021hem_03fg-like,Liu_2026_superchandra_2024igg_TARDIS}}. These events exhibit early spectra characterized by narrow, multiple lines of IMEs, particularly silicon and sulfur, with weak or absent \FeIII\ lines, a notable lack of high-velocity \CaII\ features, and strong \CII\ \llamb\ 6578, 6582 features that persist for weeks after maximum light. They also feature slow ejecta velocities along with a generally low ionization state that persists into the nebular phase \citep{ashall_2021_2003fg_superchandra}. (For a comparison see Figs.\ref{spect_comp_early} and  \ref{spect_comp_late}).
Photometrically, these events stand apart from other SN subtypes. Although their light curves are broad, they lack the secondary near-infrared (NIR) maximum typically observed around +10 days in the $H$ band of normal SNe\,Ia. Their $i$-band peak occurs after the $B$-band maximum, and they also show no distinct $H$-band break, suggesting that the synthesized \Nifs\ is buried within a dense shell \citep{ashall_2021_2003fg_superchandra}. Some of them exhibit significant bolometric dimming a few months after $B$-band maximum, causing their light curves to deviate from expected radioactive decay paths. Notable examples include SN\,2012dn \citep{taubenberger_2011dn}, SN\,2009dc \citep{hachinger_2009dc}, LSQ14fmg \citep{Hsiao_2020_LSQ14fmg}. Additionally, they often display unusually blue early colors and high UV luminosities \citep{brown_2014_distance_to_12dn}.
 While sharing these traits, their luminosities vary widely ($M_B \approx -19$ to $-21$). Some of the most luminous examples, such as SN\,2003fg \citep[M$_B = -20.18 \pm$ 0.05;][]{03-fg_original_paper}, SN\,2009dc \citep[M$_B = -20.09 \pm$ 0.17][]{yamanaka_2012dn_optical_and_NIR, taubenberger_2011_2009dc}), and SN\,2007if, \citep[M$_B = -20.54 \pm$ 0.04][]{2007if_original_discovery_paper,Scalzo_2010_dd_merger_superchandra,Taubenberger_2013}, reach luminosities intermediate between the 91T-like subclass \citep{obrien2023_1991TlikeSNe} and the superluminous SNe (SLSNe) \citep{Moriya_SLSNe}.
    These high luminosities suggest \chris{ejecta} masses exceeding the Chandrasekhar limit, ($M_{\text{Ch}}$), earning them the "Super-Chandrasekhar (SC)" designation. However, several 03fg-like objects, including ASASSN-15hy \citep{Lu_ASASSN-15hy}, SN\,2006gz \citep{hicken_2006gz}, and SN\,2012dn (this study), exhibit luminosities similar to normal SNe\,Ia. Despite their standard peak magnitudes, they retain the class's defining spectroscopic features, making them vital for probing the diversity of the most luminous SNe\,Ia.

Most intriguingly, SN\,2012dn is one of the very few SNe\,Ia known to exhibit nebular [\OI] emission. Within the 03fg-like subgroup, this feature has also been observed in SN\,2021zny \citep{Dimitriadis_2023_2021zny_OI_emission_DDmerger} and SN\,2022pul \citep{Siebert_2024}. Other examples of [\OI]-emitting SNe\,Ia include SN\,2010lp \citep{Taubenberger_2010lp}, ASASSN-20jq \citep{BOSE_2025_ASASSN-20jq_o_emission_underlum} and iPTF14atg \citep{KROMER_2016_iPTF14atg_O_emission}.

\chris{Proposed scenarios for 03fg-like SNe can be broadly described as: (i) explosions of rotating or magnetized super-Chandrasekhar white dwarfs \citep{YOON_2005_Rotating_WD_uppermasslimit}; (ii) interaction with a dense, H-poor circumstellar medium, where an additional luminosity component arises from ejecta--envelope interaction \citep{Hoeflich_Khokhlov_1996, hachinger_2009dc, Noebauer_2016__denseC_O_interacion_superchandra} (see also the core-degenerate scenario; \citealp{Sparks_stecher_CD_scenario_1974, Livio_riess_2003, Hsiao_2020, ashall_2021_2003fg_superchandra}); and (iii) the violent merger of two C--O white dwarfs whose combined mass exceeds $M_{\text{Ch}}$ \citep{Howell_2006, Scalzo_2010_dd_merger_superchandra, pakmoe_2010_WDmergers_equalmass}.}

Mapping ejecta stratification has proven essential for unraveling the spectroscopic diversity of these events \citep{Stehle2005, Mazzali2008, Sasdelli2014, Ashall2016, Aouad_99aa, Aouad_iPTF16abc}. This approach reveals potential progenitor-explosion channels and helps determine the reliability of specific subclasses as cosmological distance indicators. Furthermore, it allows for a clearer understanding of the underlying physics of these objects, such as the energetics of the explosion and the extent of nuclear burning.

The paper is organized as follows: Section \ref{data} presents the observational data, and Section \ref{modelling techniques} describes the modeling code and techniques employed. Our spectroscopic results are detailed across three sections: Section \ref{photopheric_phase} covers the photospheric phase, Section \ref{nebular_phase} examines the nebular phase, and Section \ref{abundancediscussion} presents the resulting chemical stratification. In Section \ref{bolometriclightcurvesection}, we generate a bolometric light curve based on these abundance results to test their consistency. Finally, Section \ref{discussion} evaluates potential progenitor and explosion channels, followed by our summary in Section \ref{conclusions}.

\vspace{-10pt}

\begin{figure}
\includegraphics[trim={4 0 25 12},clip,width=0.49\textwidth]{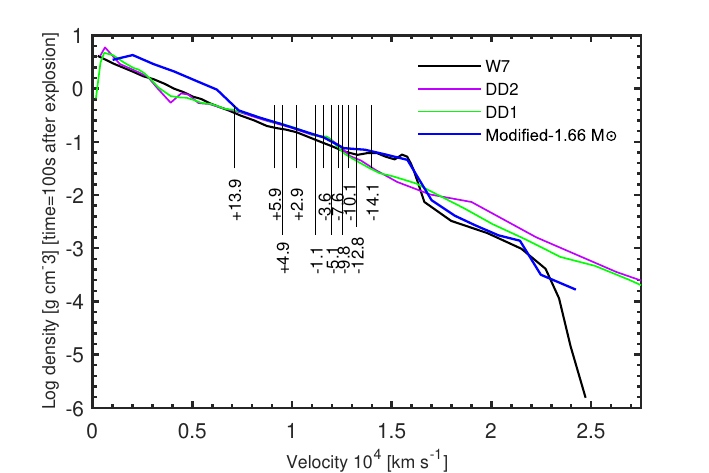}
 \caption{Density profiles for the W7, DD1, DD2 and DD3 models, all scaled to an epoch of 100\,s. The blue line shows our adopted density structure, modified to a total ejecta mass of $1.66\,M_{\odot}$. Vertical lines indicate the photospheric velocities of the synthetic spectra at their respective epochs relative to $B$-band maximum light.} 
\label{densityprofiles}
\end{figure}

\section{DATA}
\label{data}
\subsection{Host Galaxy, Distance, and Reddening}
SN\,2012dn was discovered on 2012 July 8.52 UT in the SA(s)cd spiral galaxy ESO462-016 \citep{Bock_2012_discoveryof2012dn} with a redshift of 0.010177 \citep{Springob_2005_redshift_12dn}.
Several distance moduli, $\mu$, for the host galaxy have been reported. Based on the galaxy's recessional velocity and using $H_0 = 72$ \Kms\ Mpc$^{-1}$, \citet{chakdrahari_2012dn} and \citet{brown_2014_distance_to_12dn} estimate $\mu = 33.15 \pm 0.15$ and $\mu = 33.32 \pm 0.20$ mag respectively.  Tully-Fisher measurements vary, with values of $\mu = 33.15 \pm 0.52$ mag \citep{springob_distance_to_12dn} and $\mu = 33.19 \pm 0.69$ mag \citep{lagattuta_213_distance_to_12dn}. \citet{parrent_2012dn_spectra} adopts $\mu = 33.15 \pm 0.15$ mag, \citet{taubenberger_2011dn} uses $\mu = 33.26 \pm 0.20$ mag, while \citet{yamanaka_2012dn_optical_and_NIR} adopts $\mu = 33.15$ mag, based on the work of \citet{distNGC2595Theureau34.07}.
We adopt $\mu = 33.26 \pm 0.20$ mag and a total reddening of $E(B-V) = 0.09$ (comprising $0.05$ from the Milky Way and $0.04$ from the host), following \citet{taubenberger_2011dn}. This extinction is consistent with \citet{parrent_2012dn_spectra} and \citet{yamanaka_2012dn_optical_and_NIR} (who use 0.10), though lower than the 0.18 assumed by \citet{chakdrahari_2012dn}. Reported \Dm\ values also vary between 0.92 \citep{chakdrahari_2012dn, yamanaka_2012dn_optical_and_NIR}, 0.97 \citep{taubenberger_2011dn}, and 1.08 \citep{parrent_2012dn_spectra}.

\begin{figure}
\includegraphics[trim={30 30 30 30
},clip,width=0.45\textwidth]{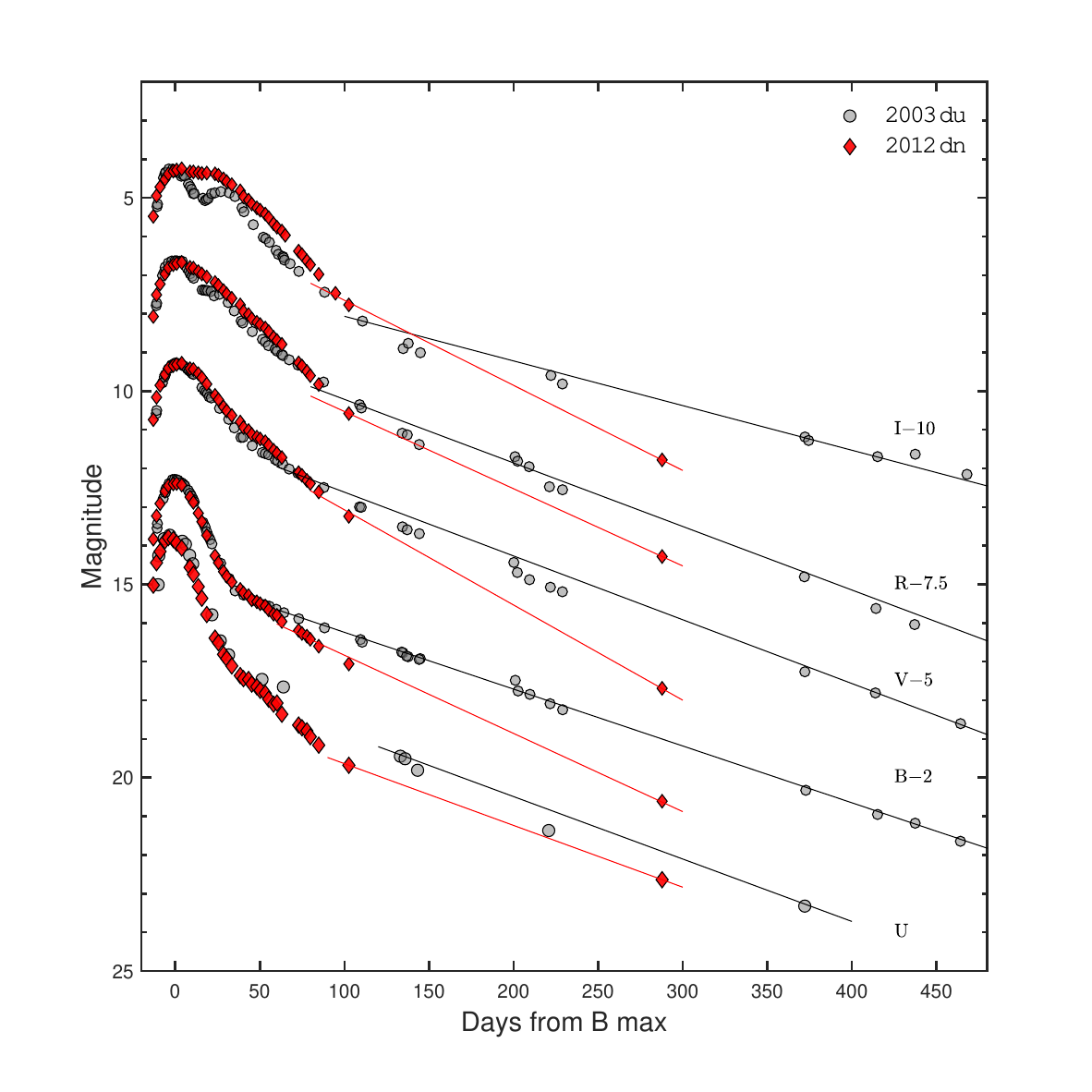}
 \caption{\textit{UBVRI} light curves or SN\,2012dn compared to the ones of SN\,2003du \citep{stanishev2003du} normalized at peak. The evolution of SN\,2012dn exhibits a notable wavelength-independent (grey) extinction that persists across all bands through the late phases.}  
\label{UBVRI_12dn_03du}
\end{figure}

\subsection{Spectra}

\begin{table}
\setlength{\tabcolsep}{2pt}

\caption{Spectra of SN\,2012dn and modelling parameters.}
\label{tab1}

\scalebox{0.95}{
\hskip-0.5cm\begin{tabular}{clcccccc}

\hline
   UT Date   &  JD$^a$ & Epoch$^b$ & Telescope/Instr. & log\,$L$ & $v$ & $T_\mathrm{ph}$ \\
    2012      &          &   (days)         & &[\Lsun] & (\kms) & (K) \\
  \hline
 10/07$^c$ & 118.5 & $-$14.1  & Gemini-S / GMOS-S & 9.10 & 14000 & 15100 \\
 11/07$^d$ & 119.8 & $-$12.8  & Lick-3m / KAST & 9.190 & 13250 & 13980 \\
 14/07$^c$ & 122.5 & $-$10.1  & FLWO-1.5M / FAST & 9.470 & 12850 & 13250 \\
 15/07$^c$ & 123.5 & $-$9.8   & Gemini-S / GMOS-S  & 9.500  & 12550 & 12990\\
 16/07$^d$ & 125.0 & $-$7.6   &  Keck1 / LRIS / & 9.570 & 12350 & 12395 \\
 19/07$^c$ & 127.5 & $-$5.1   & Gemini-S / GMOS-S  & 9.600 & 12000 & 11398 \\
 20/07$^e$ & 129.0 & $-$3.6   & ANU-2.3m / WiFes   & 9.630 & 11600 & 11285 \\
 23/07$^c$ & 131.5 & $-$1.1   & Gemini-N / GMOS  & 9.670 & 11200 & 10800 \\
 27/07$^c$ & 135.5 & $$+$$2.9 &  Gemini-N / GMOS  & 9.600 & 10200 & 9875 \\
 29/07$^c$ & 137.5 & $$+$$4.9 & FTS / FLOYD-S  &  9.560 &  9500 & 9486 \\
 30/07$^c$ & 138.5 & $$+$$5.9 & Gemini-N / GMOS  &  9.550 &  9100 & 9438 \\
 04/08$^c$ & 146.5 & $$+$$13.9 & FTS / FLOYD-S  &  9.440 &  7100 & 8449\\
\hline
\multicolumn{7}{c}{Nebular Phase$^f$} \\
\hline
- & 423.3  & $+$290.9 &  ESO 8.2m/FORS2-ISAAC  &  -- &  -- & -- \\
\hline
\end{tabular}}
\small
$^a$ JD$-$2,456,000 -- $^b$ Rest-frame time, since $B$ maximum \\
$^c$~\citep{parrent_2012dn_spectra},
$^d$~\citep{stahl_12dn_spectra_reference},
$^e$~\citep{childress2016_12dn_spectra_ref},
$^f$~\citep{taubenberger_2011dn}\\

\label{modeling_parameters}
\end{table}

The spectra used in this study were retrieved from the Weizmann Interactive Supernova Data Repository (WISeREP; \citealt{wiserep}) and are listed in Table~\ref{tab1}. \referee{The original spectroscopic observations and data reductions were published by \citet{parrent_2012dn_spectra, stahl_12dn_spectra_reference, childress2016_12dn_spectra_ref} and \citet{taubenberger_2011dn}.}
Following  \citet{Aouad_99aa, Aouad_iPTF16abc}, flux calibration in the $U, B, V, R, I$ was performed  by multiplying the spectra with a low-order smoothed spline using the \chris{photometric} data of \citet{taubenberger_2011dn}. 
\vspace{-30pt}
\subsection{Photometric Evolution and Late-Time Dimming}
The photometry of SN\,2012dn \citep{taubenberger_2011dn}, reveals a distinct light-curve evolution. While the peak luminosity is comparable to normal SNe\,Ia, the post-maximum decline is highly unusual; starting $\sim$60 days post-maximum, the light curves fade significantly faster than standard events. A comparison with the normal SN\,2003du (Fig. \ref{UBVRI_12dn_03du}) highlights a nearly wavelength-independent (grey) luminosity deficit across all \textit{UBVRI} bands through day 300 (quantified in Appendix \ref{grey_dimming}; see also Sec. \ref{nebular_phase}.) Similar late-time behavior has been observed in other 03fg-like candidates such as SN\,2006gz \citep{Maeda_2006gz_latephases}, LSQ14fmg \citep{Hsiao_2020_LSQ14fmg}, and SN\,2009dc \citep{taubenberger_2011_2009dc}, though the onset in SN\,2012dn is notably earlier.

\referee{The physical origin of the late-time luminosity deficit remains uncertain. \citet{taubenberger_2011dn} examined several possible explanations, namely enhanced $\gamma$-ray or positron escape, clumping, an infrared catastrophe (IRC), and dust formation. Early positron escape appears unlikely, as reproducing the observed deficit would require unrealistically low ejecta masses \citep{taubenberger_2011dn}, Moreover, recent observations favour efficient positron trapping in normal Type Ia supernovae at epochs far later than those considered here \citep{Kerzendorf_2014_late_time_photom_2011fe,Kerzendorf_2017_extreme_late_time_photom_2011fe,Graur_2016_late_time_photom_2012cg,Shappee_2017,Leloudas_2009_no_poisitronss_escape_before800days,Fransson_Jerkstrand_2015_no_positrons_escape}. As also noted by \citet{taubenberger_2011dn}, strong clumping does not resolve the discrepancy, since reducing the positron optical depth would simultaneously reduce the $\gamma$-ray optical depth, producing an early decline incompatible with the relatively slow light-curve evolution of SN\,2012dn.}

\referee{Likewise, an IRC appears unlikely. At $\sim300$ days, the nebular spectrum remains dominated by thermally excited [\FeII] emission, while the detection of [\OI] and [\CaII] indicates that the ejecta remain in the thermal regime. More broadly, no convincing evidence for an IRC has been reported in normal SNe\,Ia \citep{Graur_2016_late_time_photom_2012cg,Kerzendorf_2014_late_time_photom_2011fe,Leloudas_2009_no_poisitronss_escape_before800days}.}

\referee{Taken together, these arguments leave dust as the most plausible explanation for the observed luminosity deficit \citep{taubenberger_2011dn,yamanaka_2012dn_optical_and_NIR,nagao_2017_IR_12dn,parrent_2012dn_spectra,chakdrahari_2012dn}. 
In particular, the interpretation in terms of newly formed dust within the ejecta \citep{taubenberger_2011dn}, is supported by theoretical models predicting dust condensation in Type Ia ejecta at relatively early epochs ($\sim100$--300 days; \citealt{Nozawa2011_dust_in_Ias}) and by the delayed onset of the luminosity deficit after maximum light, which is more naturally explained by dust forming within the expanding ejecta than by pre-existing dust.
However, the nearly wavelength-independent fading, accompanied by little or no colour evolution, disfavors conventional interstellar-like dust, instead favouring either unusually large grains \citep{aguirre_99_greydust} or optically thick dust clumps \citep{taubenberger_2011dn}. In the latter scenario, the clumps shield newly formed grains from the intense radiation field, relaxing the requirement for rapid growth to unusually large grain sizes while naturally producing the observed near-grey attenuation. Nevertheless, without near- and far-infrared observations at the nebular epoch, the origin and physical properties of the dust remain difficult to constrain observationally.}

\section{PHOTOSPHERIC PHASE MODELS}
\label{photopheric_phase}

\subsection{Modelling Technique}
\label{modelling techniques}

Radiative-transfer calculations are performed using the Monte-Carlo code developed by \citet{Mazzali1993} and extended later by \citet{Lucy1999} and \citet{Mazzali2000}. The code is described extensively in \citet{Stehle2005, hachinger_2011_phdthesis, Hachinger2017blueshiftedNA, ashall2020error}. Assuming homologous expansion, a sharp photosphere is imposed at velocity $v_{\rm ph}$ (where $R_{\rm ph} = v_{\rm ph} \cdot t$), emitting a blackbody continuum with luminosity $L = 4\pi (v_{\rm ph}t)^{2}\,\sigma T_{\rm ph}^{4}$. This luminosity is discretized into photon packets that undergo Thomson scattering and Sobolev line interactions within the atmosphere with bound-bound emissivity treated via a branching scheme \citep{Mazzali2000}. Through estimators, the code calculates, in each shell, the mean intensity, $J = \int_{0}^{\infty} J_{\nu}\, d\nu$ and the frequency moment, $\langle \nu \rangle = \frac{\int_{0}^{\infty} \nu J_{\nu}\, d\nu}{\int_{0}^{\infty} J_{\nu}\, d\nu}$.
 These define the radiation temperature $T_R$ and dilution factor $W$, via: $\frac{h\langle \nu \rangle}{kT_R} = 3.832 \quad \text{and} \quad W(r) = \frac{J(r)}{B(T_R(r))}$. Using $T_R$, $W$, and $T_e \simeq 0.9\,T_R$, the code updates ionization and excitation via the modified nebular approximation \citep{Mazzali1993, Mazzali2000}. Radiative equilibrium is enforced \citep{Lucy1999}. Packet propagation and level populations are iterated until convergence. 
 The emergent spectrum is then obtained from a formal solution of the radiative-transfer equation using the converged source function. To initiate the modeling, the code requires a fixed density profile, the time since explosion $t$, a bolometric luminosity $L_{\rm bol}$, the photospheric velocity $v_{\rm ph}$, and the chemical abundances for every shell. These parameters are manually optimized to match the observed spectra. As the ejecta expand and dilute over time, the photosphere recedes in velocity space, revealing deeper layers.

\begin{figure*}

\includegraphics[trim={0 0 61 0},clip,width=0.8\textwidth]{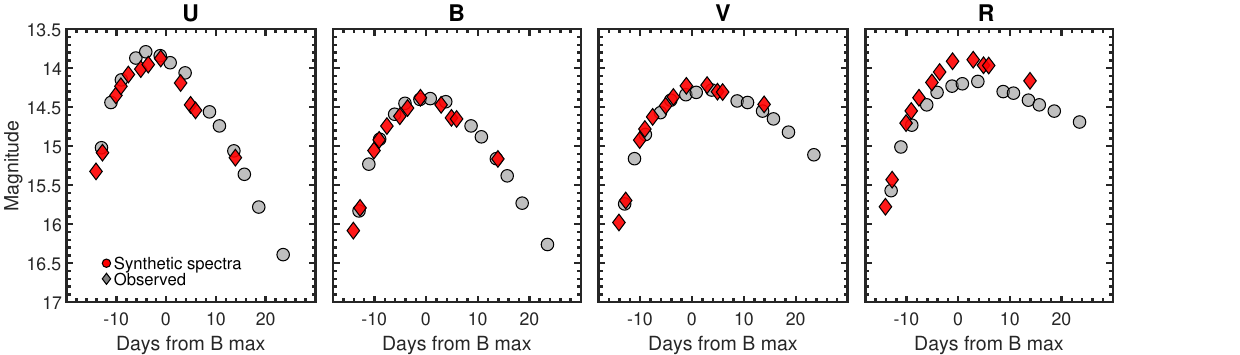}
 \caption{Light curves from the synthetic spectra compared to the light curves from observation. Photometry used from \citet{taubenberger2017}.}  
\label{UBVR}
\end{figure*}

We examined 12 spectra ranging from day $-14$ to $+14$ relative to $B$-band maximum. We employed two distinct density profiles, illustrated in Figure~\ref{densityprofiles}, corresponding to two different explosion models: a fast deflagration (W7; \citealp{nomoto1984}) and a delayed detonation model (DD2; \citealp{iwamoto1999}). These profiles have been consistently applied across previous studies in this series, ensuring a standardized comparison across different supernovae. Our synthetic spectra provide an exceptional fit to the observed features and flux levels across the entire wavelength range (See Fig.\ref{UBVR}). Input parameters for the modeling are summarized in Table \ref{modeling_parameters}.

\subsection{The pre-maximum spectra}

\begin{figure*}
 
\captionsetup[subfigure]{labelformat=empty}
	\centering

 \begin{subfigure}{1\textwidth}
		\includegraphics[trim={35 10 52 14},clip,width=0.78\textwidth]{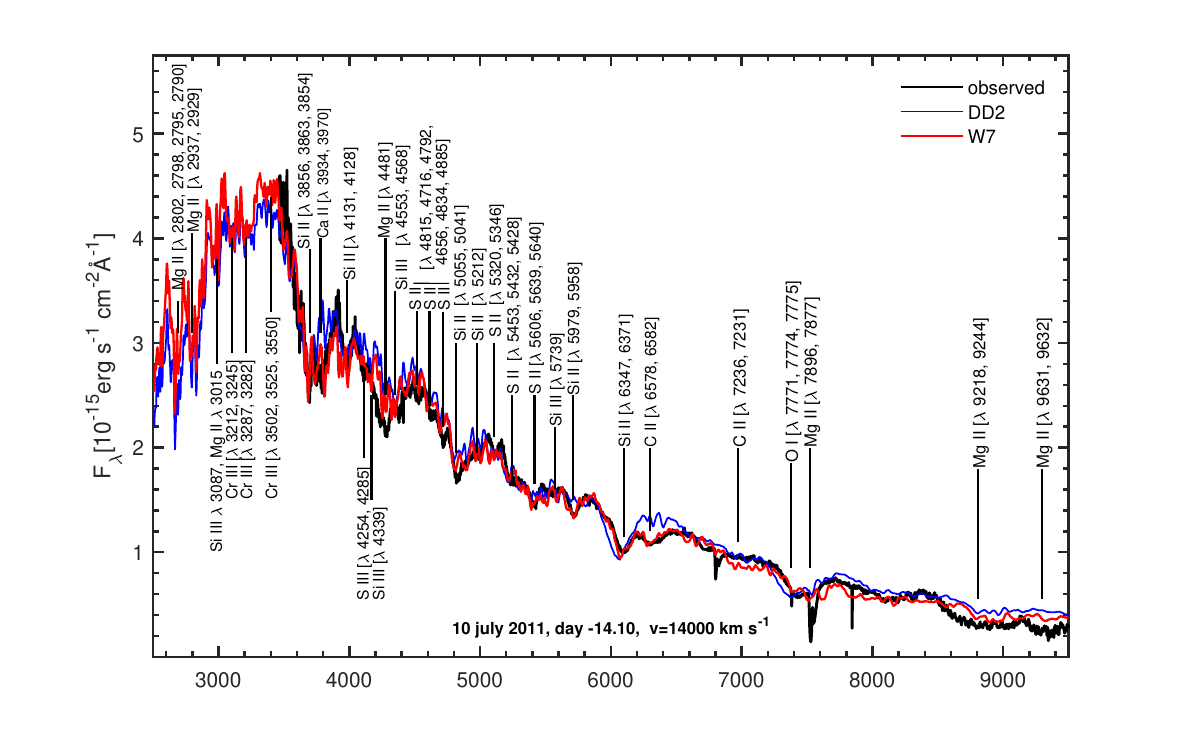}
        \centering
		\caption{ }
       
	\end{subfigure}
 \vspace{-35pt}

 \begin{subfigure}{1\textwidth}
		\includegraphics[trim={35 1 52 14},clip,width=0.78\textwidth]{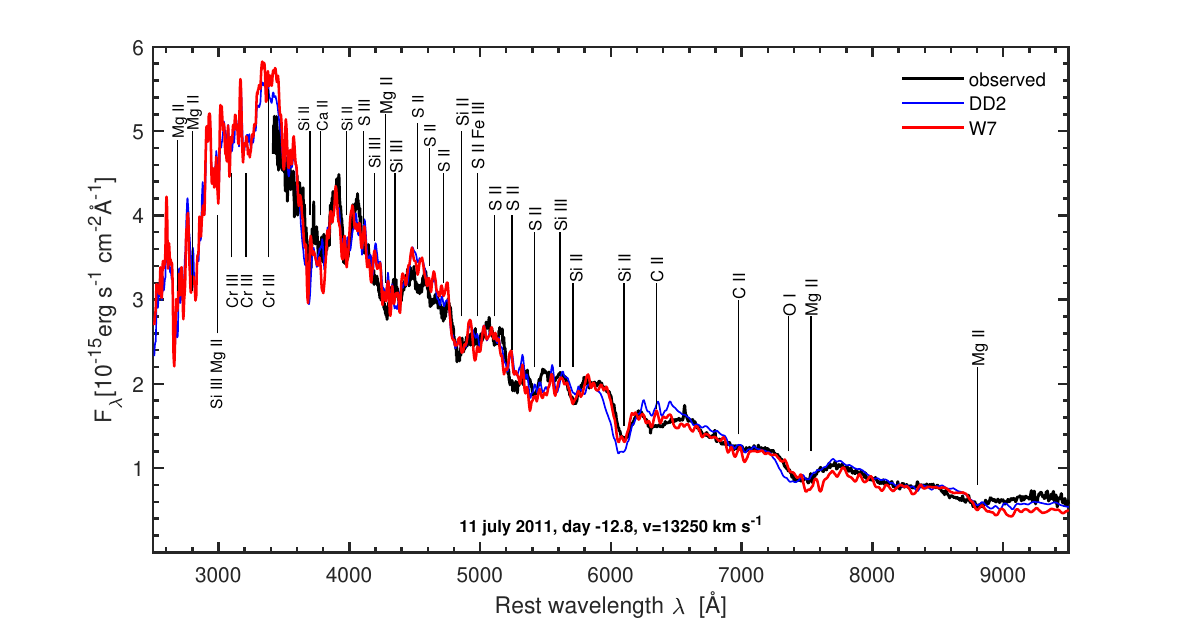}
        \centering
		\caption{ }
	\end{subfigure}

\vspace{-20pt}
 \caption{ Early-time spectra of SN\,2012dn 
 overlaid with the synthetic spectra computed using the W7 and the DD2 density profiles depicted in red and blue, respectively. }
 \label{dayearly1}
\end{figure*}

\begin{figure*}
 \label{spect-10-9}
\captionsetup[subfigure]{labelformat=empty}
	\centering

 \begin{subfigure}{1\textwidth}
		\includegraphics[trim={35 5 52 14},clip,width=0.78\textwidth]{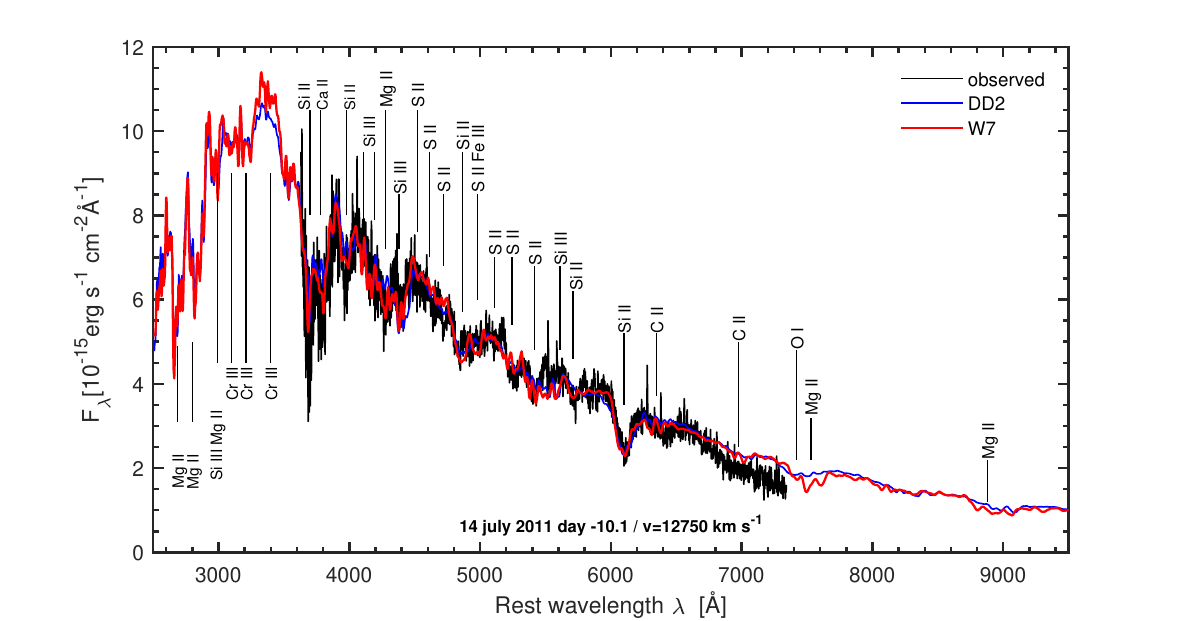}
        \centering
		\caption{ }
	\end{subfigure}
 \vspace{-35pt}

 \begin{subfigure}{1\textwidth}
		\includegraphics[trim={35 1 52 14},clip,width=0.78\textwidth]{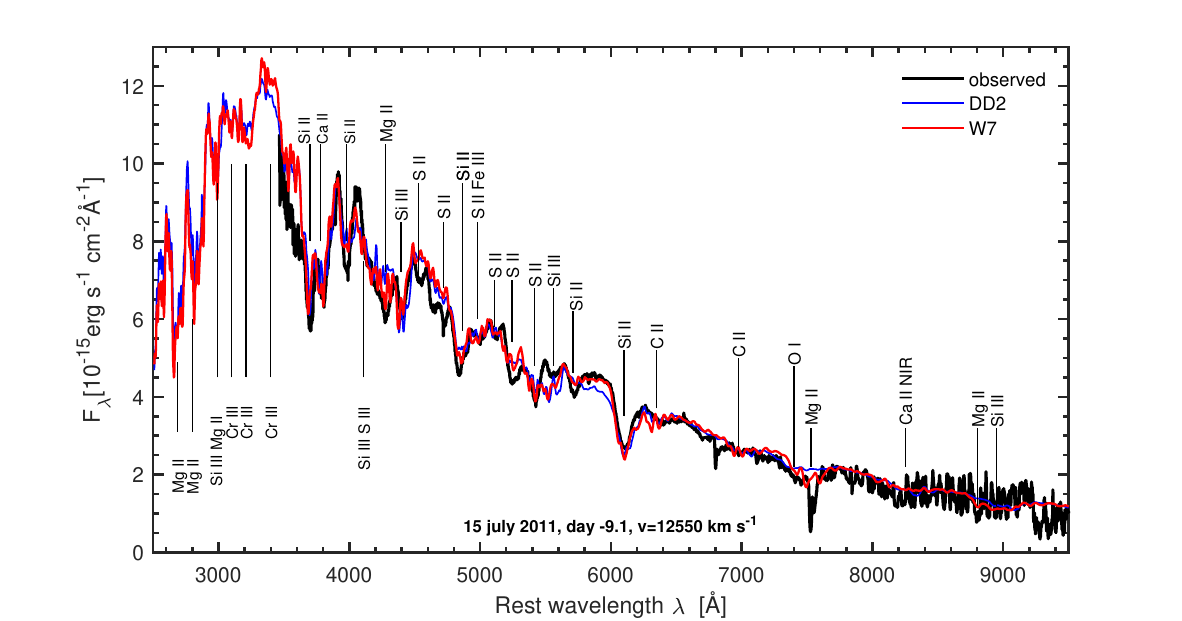}
        \centering
		\caption{ }
	\end{subfigure}

\vspace{-20pt}
 \caption{ Early-time spectra of SN\,2012dn 
 overlaid with the synthetic spectra computed using the W7 and the DD2 density profiles depicted in red and blue, respectively. }
\label{dayearly2}
\end{figure*}

\begin{figure*}
 \label{spect-7-5}
\captionsetup[subfigure]{labelformat=empty}
	\centering

 \begin{subfigure}{1\textwidth}
		\includegraphics[trim={35 5 52 14},clip,width=0.78\textwidth]{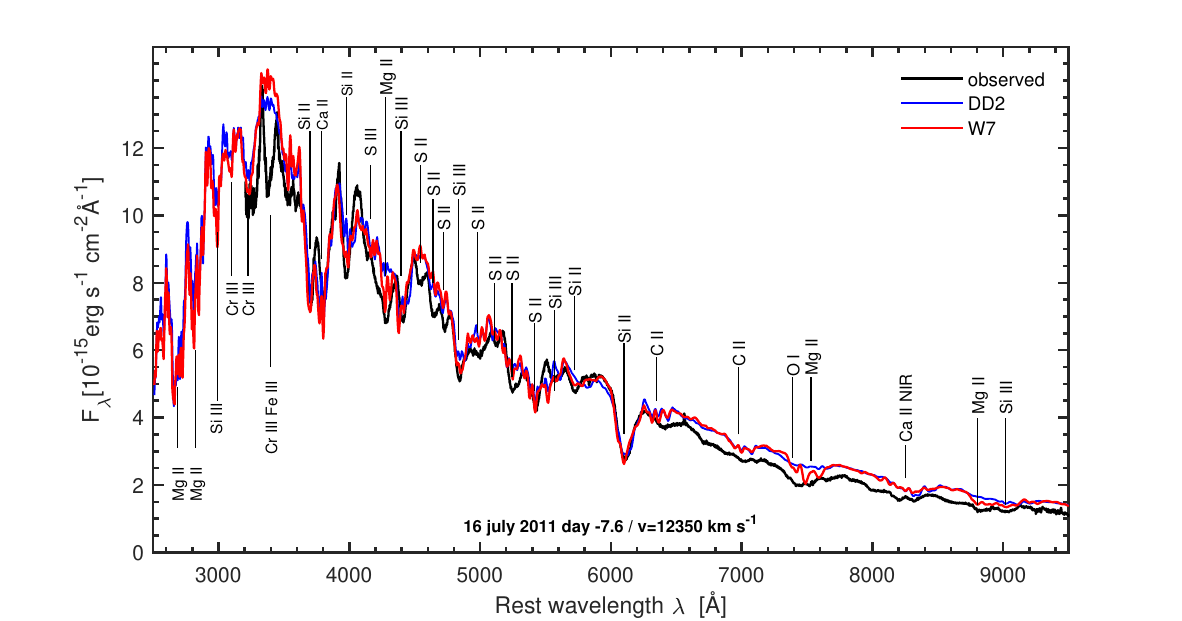}
        \centering
		\caption{ }
        
	\end{subfigure}
 \vspace{-35pt}

 \begin{subfigure}{1\textwidth}
		\includegraphics[trim={35 1 52 14},clip,width=0.78\textwidth]{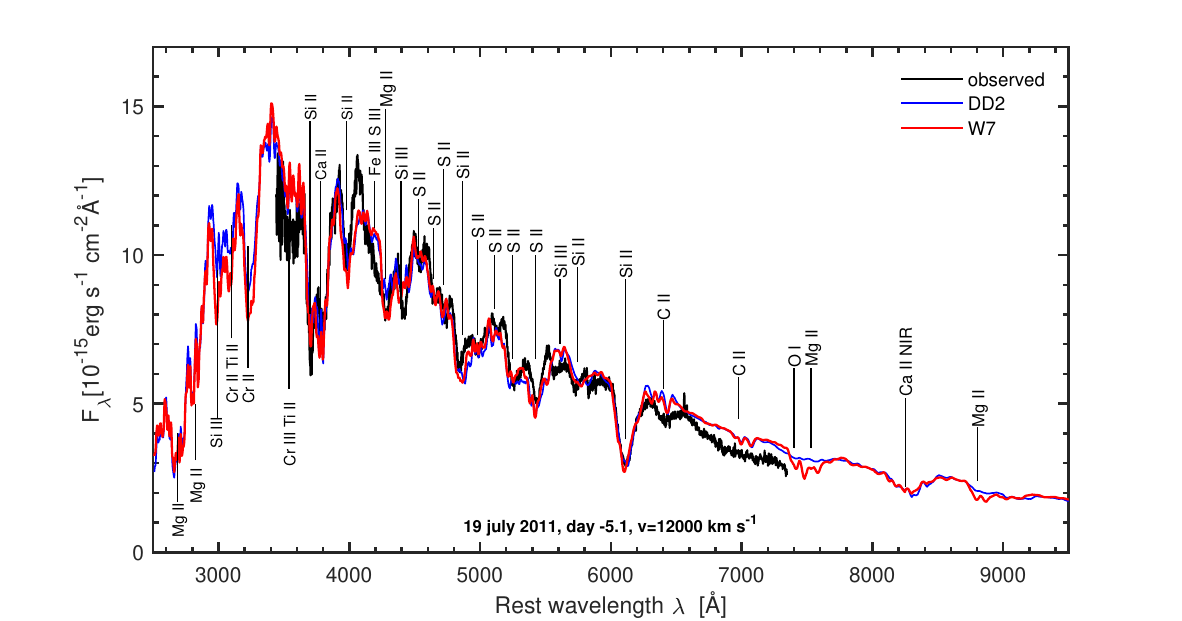}
        \centering
		\caption{ }
	\end{subfigure}

\vspace{-20pt}
 \caption{ Early-time spectra of SN\,2012dn 
 overlaid with the synthetic spectra computed using the W7 and the DD2 density profiles depicted in red and blue, respectively. 
	}
 \label{dayearly3}
\end{figure*}

Figures \ref{dayearly1}–\ref{dayearly3} present synthetic spectra from day $-$14.1 to $-$5.1. At these early epochs, relatively high photospheric velocities are required compared to SNe\,2009dc, 1999aa, iPTF16abc, and 2003du at similar phases. Higher velocities imply lower temperatures for a given luminosity, which is necessary to reproduce the observed low ionization.


\begin{figure*}
\includegraphics[trim={5 2 10 20},clip,width=0.8\textwidth]{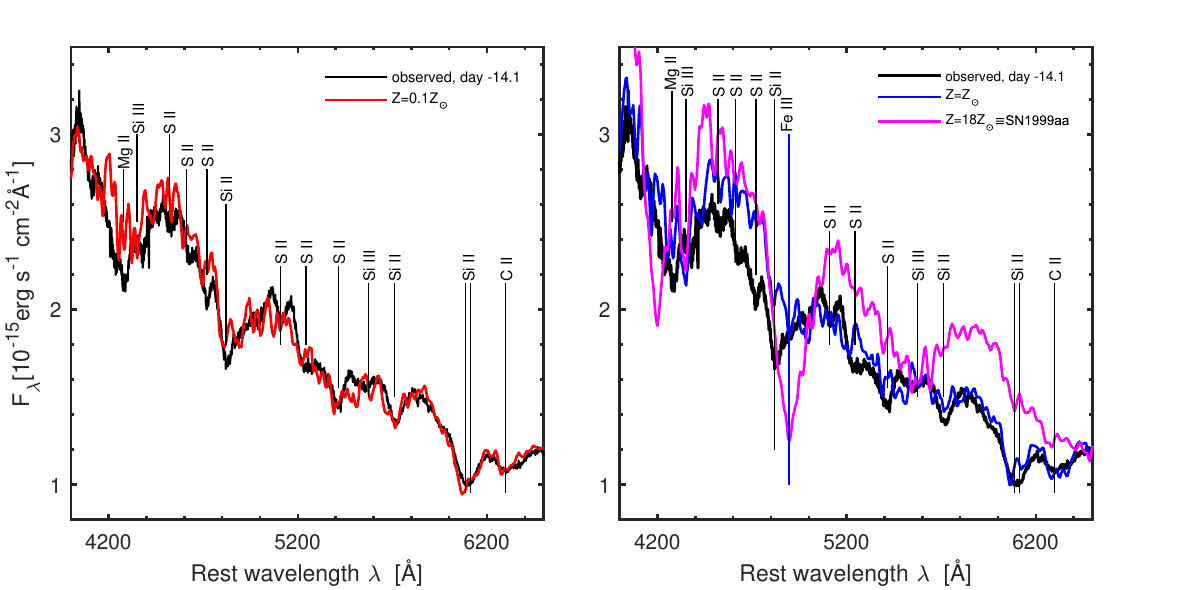}
 \caption{Probing outer-layer iron: Left panel shows the day $-$14 spectrum with a sub-solar model ($Z = 0.1\,Z_{\odot}$) at $v > 15,000$ \kms. Right panel compares solar (blue) and super-solar (purple; matching SN 1999aa) abundances. Solar and super-solar abundances fail to replicate the observed spectral features. 
 }

\label{probing_Fe}
\end{figure*}

\textit{Fe-group elements:} Unlike normal or 91T-like supernovae, the early spectra of SN\,2012dn and other 03fg-like events (\eg\ SN\,2006gz; \citealt{hicken_2006gz}) lack visible \FeIII\ lines \citep{ashall_2021_2003fg_superchandra}. The 4250 \AA\ feature is dominated by \MgII, as in normal SNe\,Ia rather than by \FeIII\ as in 91T-like events. Likewise, the shallow 5000 \AA\ feature is driven primarily by \SiII\ $\lambda\lambda$5041, 5056 instead of the strong \FeIII\ absorption typically seen in both normal and 91T-like SNe.

 For $v > 15000$ \kms, no iron is required in our models, with the maximum allowable iron mass fraction being less than 0.0001. This is significantly lower than the solar value \citep[$X$(Fe$_\odot$) = 0.001,][]{asplund-2009-solar-abundance}.
 To further constrain this, we computed synthetic spectra with metallicities of $Z = Z_\odot$ and $Z = 18\,Z_\odot$, as shown in Fig.~\ref{probing_Fe}. Increasing the iron content in the outermost layers overestimates the early-time flux and introduces noticeable discrepancies in IME features. Specifically, the \SiII\ $\lambda\lambda$5041, 5056 lines are better reproduced with sub-solar abundances, whereas solar levels introduce unwanted \FeIII\ influence near 4900\,\AA. Furthermore, higher iron abundances enhance backscattering, which heats the photosphere and shifts the ionization balance from Si II toward \SiIII. This results in the strengthening of the \SiIII $\lambda\lambda$ 4552, 4567, 4574 multiplet near 4300 \AA, which is absent in the spectra of SN\,2012dn.

\begin{figure}
\includegraphics[trim={2 0 24 15},clip,width=0.354\textwidth]{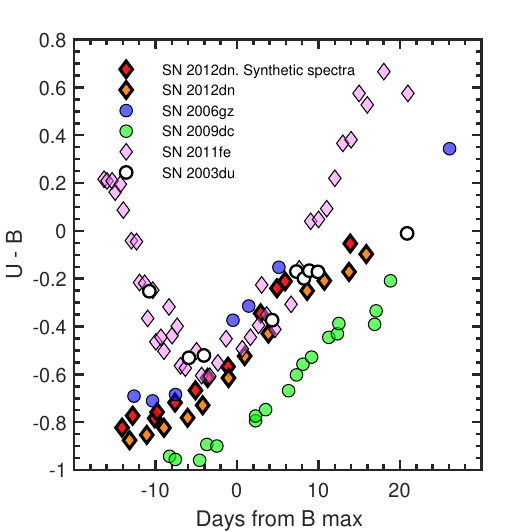}
 \caption{$U$--$B$ color evolution of SN\,2012dn compared with synthetic spectra and other SNe \citep{taubenberger_2011dn}. The 03fg-like events exhibit notably blue pre-maximum colors, likely due to a lack of Fe-group elements reducing UV line blanketing \citep{mazzali_2014_2011fe}. This contributes to the distinct early color evolution observed in this subclass.
 }  
 
\label{U-B}
\end{figure}

\cha{The low iron abundance aligns with studies of the host galaxy of SN\,2012dn, which consistently report sub-solar metallicities ranging from $\sim 0.4$ to $0.8\,Z_{\odot}$ \citep{parrent_2012dn_spectra, taubenberger_2011dn, childress2016_12dn_spectra_ref}. Furthermore, the lack of iron-group elements reduces UV line blanketing \citep{mazzali_2014_2011fe}, explaining both the extreme UV brightness and the unusually blue pre-maximum $(U-B)$ colours observed in SN\,2012dn, SN\,2009dc, and other 03fg-like events \citep{brown_2014_distance_to_12dn, Lu_ASASSN-15hy, silverman_2013_CSM_Interacting_SNe, taubenberger_2011dn}. Our synthetic $(U-B)$ colour evolution reproduces the observations remarkably well (Fig. \ref{U-B}), supporting the consensus that this subclass preferentially occurs in  low-metallicity progenitors \citep{childress_2011_2007ifmetalpoorgalaxy, ashall_2021_2003fg_superchandra}.}

Going inwards, the iron mass fraction increases 
to 0.1 at $12000 < v < 12,300$ \kms.  This stable iron (\ie \Feff), which could not have originated from \Nifs\ decay at such an early stage, must be the result of explosive nucleosynthesis \citep{Tanaka2011, Aouad_99aa}. It is necessary to absorb UV flux and reprocess it to longer wavelengths.

Typically, \Nifs\ is identified by a prominent \CoIII\ line $\lambda\lambda$ 3287, 3305, observed near 3200 \AA. Unfortunately, the spectra available for SN\,2012dn do not cover this range, making it challenging to determine an accurate amount of \Nifs. 

\textit{Calcium:} Unlike normal SNe\,Ia, SN\,2012dn exhibits a weak high-velocity (HV) \CaII\ H\&K feature, a characteristic of 03fg-like SNe \citep{taubenberger_2011dn, ashall_2021_2003fg_superchandra}. In normal and transitional events, such as SN\,2003du and SN\,1999aa, the HV and photospheric (PV) \CaII\ components blend into a single deep absorption dominated by the HV contribution \citep{Aouad_99aa, Aouad_iPTF16abc, dutta_2017hpa_decomposition_of_ca_lines, childress2014_HVfeatures}. In SN\,2012dn, however, no HV component is detected, and the blue-side absorption is instead produced mainly by \SiII\ $\lambda\lambda$3854, 3856, 3863. This resembles iPTF16abc, except that the same region there is dominated by \SiIII\ and \CoIII\ rather than \SiII\ \citep{Aouad_iPTF16abc,Liu_2023wrk_iptf16abc_like}, further emphasizing the lower ionization state of SN\,2012dn relative to 91T-like events.

Unlike normal and 91T-like SNe\,Ia, our models require no Ca above $16,000$ \kms. A negligible Ca mass fraction ($5 \times 10^{-4}$) is sufficient at $15,000$--$16,000$ \kms, rising to 0.05 at $12,500$--$13,500$ \kms\ to reproduce the photospheric component. The \CaII\ NIR triplet is weak in the earliest spectra, becoming prominent only by day $-7.6$, unlike normal SNe\,Ia where it is strong from the earliest epochs, but earlier than in 91T-like events where it typically appears about a week after maximum light.


Distinguishing whether the \CaII\ line behavior stems from abundance or ionization is challenging \citep{mazzali2005HVfeatures, Aouad_99aa}. 
However, given the generally low ionization state of SN\,2012dn, the absence of high-velocity \CaII\ H\&K features is more likely an abundance effect than an ionization one.\\

\textit{Silicon, Sulphur, Magnesium:} 
The \SiII\ \lam 6355 ($\lambda\lambda$6347, 6371) and \SiII\ \lam5972 ($\lambda\lambda$5958, 5979) 
lines are clearly visible from the earliest spectrum. The accurate reproduction of their evolving depth ratio, $\mathcal{R}$ \SiII, a key ionization indicator \citep{nugent1995_Rsi, hachinger2008_RSi}, indicates that our models accurately capture the ejecta temperature and ionization structure. 
The \SiII\ \lam\ 4130 line is notably stronger than in normal and 91T-like SNe during this pre-maximum phase. \SiII\ \llamb\ 3856, 3863, 3854 create a deep absorption blueward of the \CaII\ H\&K PV feature near 3700 \AA. A trait absent in both normal and 91T-like events.

The 4850 \AA\ absorption is well reproduced by our models and is dominated by \SiII\ \llamb\ 5041, 5056, unlike in normal and 91T-like SNe\,Ia where it is produced primarily by \FeIII\ \llamb\ 5076, 5063.
Our models reproduce the \SiIII\ \llamb\ 4553, 4568 feature near 4400 \AA\ and the absorption around 5600 \AA\ blueward of \SiII\ \lam\ 5972. While \citet{taubenberger_2011_2009dc} suggested the latter, uncommon in normal and 91T-like SNe, may be a Na I D +\SiII\ blend, our models identify it as \SiIII\ \lam\ 5739. 




The characteristic W-shaped \SII\ feature near 5400 \AA, absent at this epoch in hotter 91T-like SNe, as well as the uncommon \SII\ \llamb\ 5320, 5346 feature around 5050 \AA\ and several \SII\ lines between 4500 and 4700 \AA\ are well replicated. 

The typical \MgII\ \lam\ 4481 line observed near 4200 \AA\ is well replicated. 
An \MgII\ \lam\ 7896 line blends with the \OI\ \lam\ 7771 near 7400 \AA\ broadening the line. A feature near 8800 \AA\ emerges early and persists as the spectra evolve, originating from \MgII\ \llamb\ 9218, 9244.
On day $-$ 7.6 this feature splits into two components, the redder one being mainly influenced by \SiIII\ \lam\ 9323 (See Fig. \ref{dayearly3}). 

\textit{Carbon, Oxygen:} Carbon, the signature of unburnt material, appears sporadically in SNe\,Ia \citep{folatelli_unburned_material, Parrentcarbonfeatures2011} but is ubiquitous in 03fg-like events \citep{ashall_2021_2003fg_superchandra, taubenberger_2011_2009dc}. In SN\,2012dn, both the \CII\ \lam\ 6580 and \CII\ \lam\ 7234 lines are visible early and persist as the spectra evolve. 
These carbon signatures are uncommon in 91T-like events but have been exceptionally observed in iPTF16abc and recently in SN\,2023wrk \citep{Liu_2023wrk_iptf16abc_like}. Interestingly, in these cases, the carbon features fade and then reappear post-maximum, whereas in SN\,2012dn they remain consistent down to deeper layers.
The typical \OI\ \lam\ 7771 line, near 7400 \AA, is blended with a strong \MgII\ feature on its redward side. 
\vspace{-5pt}

\begin{figure}
    \centering
    \begin{subfigure}{.25\textwidth}
        \centering
        \includegraphics[width=\textwidth]{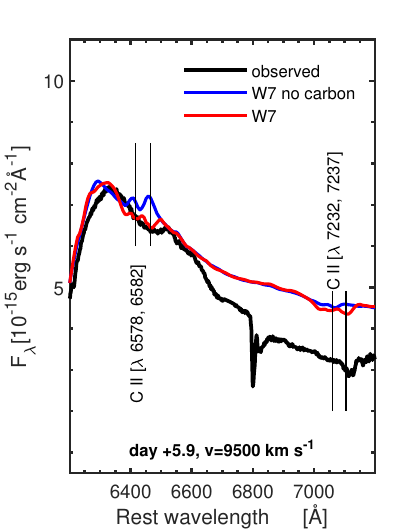}
        
    \end{subfigure}%
    \begin{subfigure}{.25\textwidth}
        \centering
        \includegraphics[width=\textwidth]{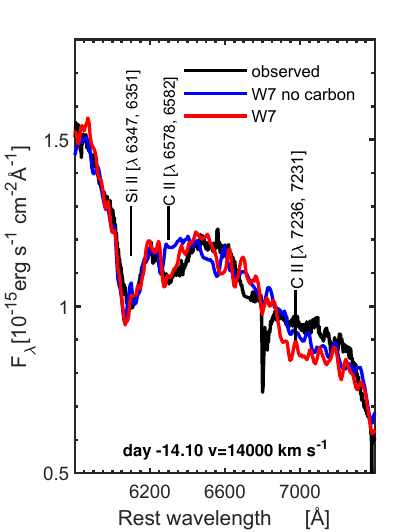}
       
    \end{subfigure}%
    
    \caption {Comparison between a synthetic model excluding carbon (blue) and a model incorporating $0.01\,M_{\odot}$ of carbon ($X(C)=0.05$) within the velocity range $9,100 < v < 10,200$ \kms\ (red) is shown in the left panel. The right panel displays the results for the outer layers, where the observed C II feature near 6,200 \AA\ is accurately reproduced by including $0.003\,M_{\odot}$ of carbon at $v > 14,000$ \kms\ (red).
    }
    \label{probing_C}
    \vspace{-11pt}
\end{figure}

\subsection{The spectra near maximum light}

\begin{figure*}
 \label{spect-4-1}
\captionsetup[subfigure]{labelformat=empty}
	\centering

 \begin{subfigure}{1\textwidth}
		\includegraphics[trim={35 5 52 14},clip,width=0.78\textwidth]{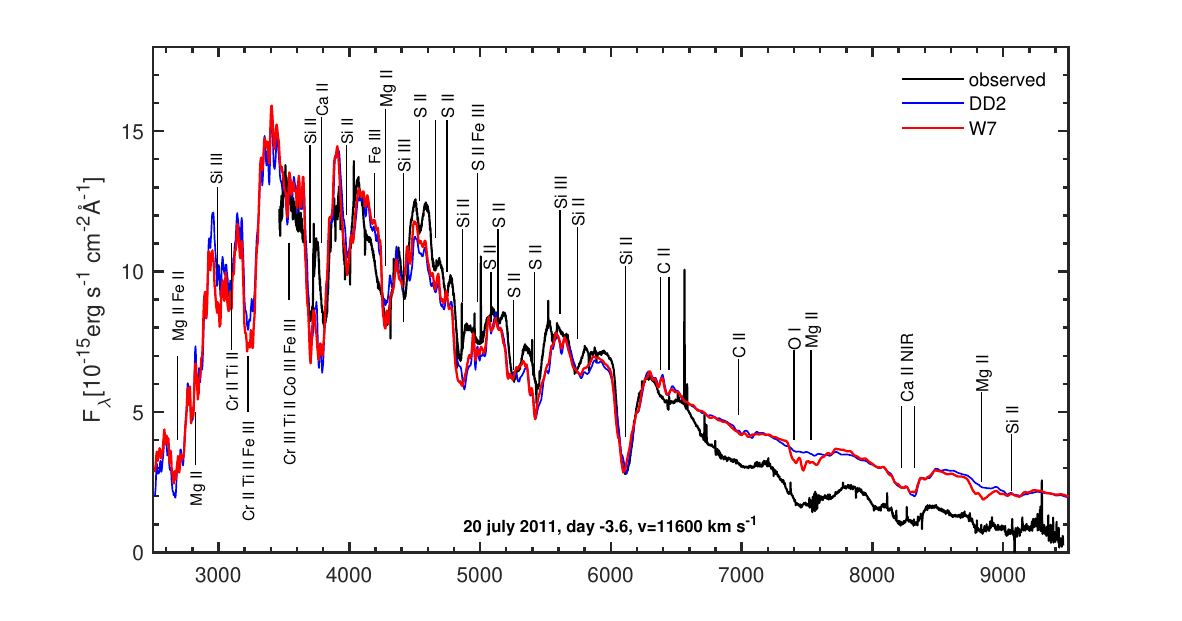}
        \centering
		\caption{ }
	\end{subfigure}
 \vspace{-35pt}

 \begin{subfigure}{1\textwidth}
		\includegraphics[trim={35 1 52 14},clip,width=0.78\textwidth]{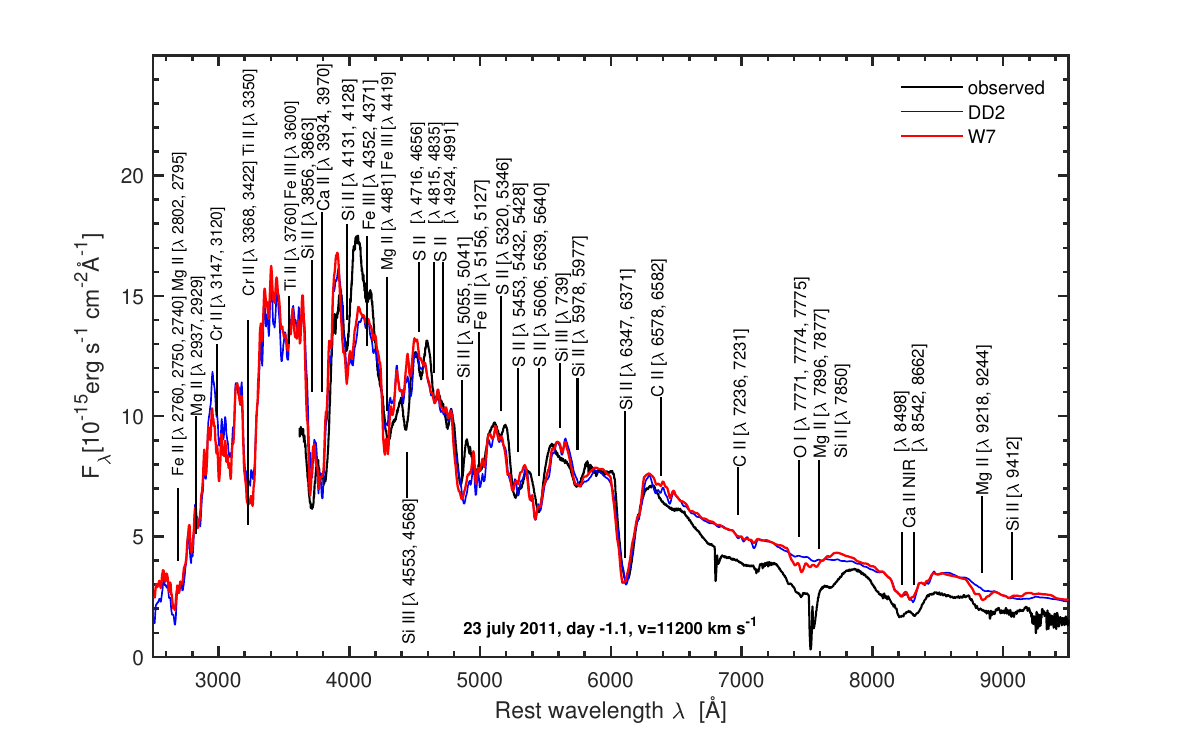}
        \centering
		\caption{ }
	\end{subfigure}

	
\vspace{-20pt}
 \caption{ Spectra of SN\,2012dn near maximum light 
 overlaid with the synthetic spectra computed using the W7 and the DD2 density profiles depicted in red and blue, respectively.
	}
 \label{daymax1}
\end{figure*}

\begin{figure*}
 \label{spect3_5}
\captionsetup[subfigure]{labelformat=empty}
	\centering

 \begin{subfigure}{1\textwidth}
 \centering
		\includegraphics[trim={35 5 52 14},clip,width=0.78\textwidth]{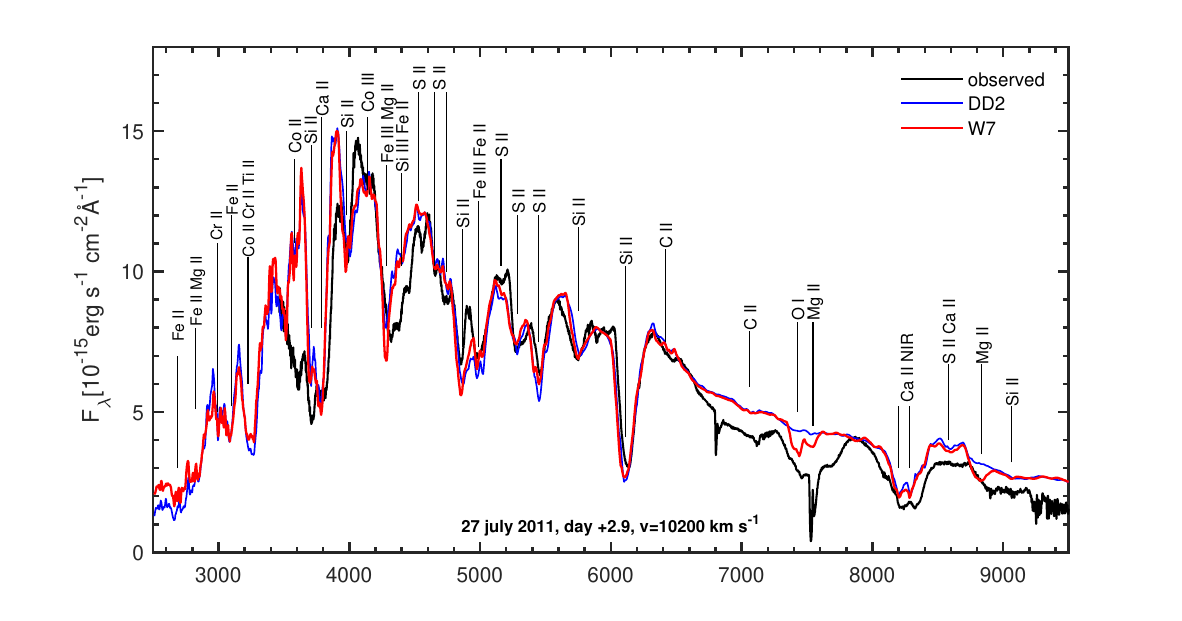}
		\caption{ }
	\end{subfigure}
 \vspace{-35pt}

 \begin{subfigure}{1\textwidth}
 \centering
		\includegraphics[trim={35 1 52 14},clip,width=0.78\textwidth]{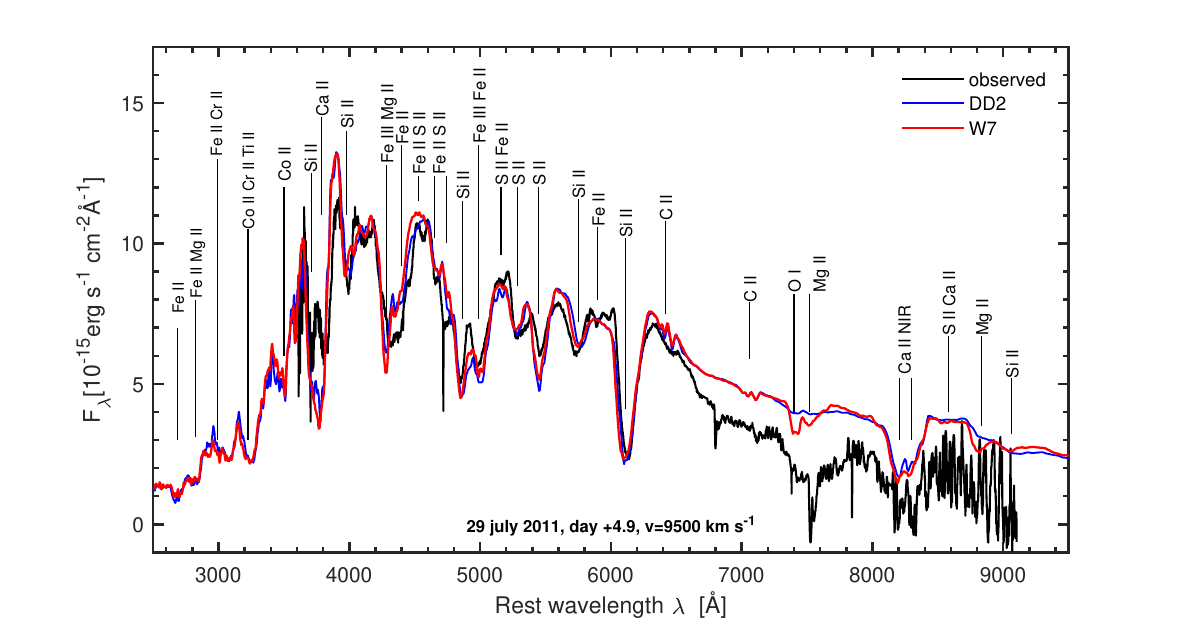}
		\caption{ }
	\end{subfigure}

	
\vspace{-20pt}
 \caption{Spectra of SN\,2012dn near maximum light 
 overlaid with the synthetic spectra computed using the W7 and the DD2 density profiles depicted in red and blue, respectively. 
}
 \label{daymax2}
\end{figure*}

Figures \ref{daymax1} and \ref{daymax2} compare synthetic and observed spectra from day $-3.6$ to $+4.9$. As the supernova approaches maximum light, its spectrum develops broader features and stronger Fe lines. Consistent with the pre-maximum analysis, the photospheric velocity remains 1,500–2,000 \kms\ higher than in SN\,2009dc \citep{hachinger_2009dc}, SN\,1999aa \citep{Aouad_99aa}, iPTF16abc \citep{Aouad_iPTF16abc}, and SN\,2003du \citep{Tanaka2011}, maintaining the low-ionization conditions required to match the observed spectra.


\textit{Fe-group elements:} The feature near 4300 \AA\ becomes more pronounced starting on day $-$1.1, driven by the emergence of \FeIII\ \llamb\ 4419, 4431, 4395. These lines, along with \MgII\ \lam\ 4481, contribute to the overall deepening of the feature as the spectra evolve. The redder component, near 4400 \AA, remains dominated by \SiIII\ \llamb\ 4553, 4568. At this epoch, the evolution of this feature closely mirrors that of both normal and 91T-like events.

\begin{figure}
\includegraphics[trim={2 2 30 18},clip,width=0.4\textwidth]{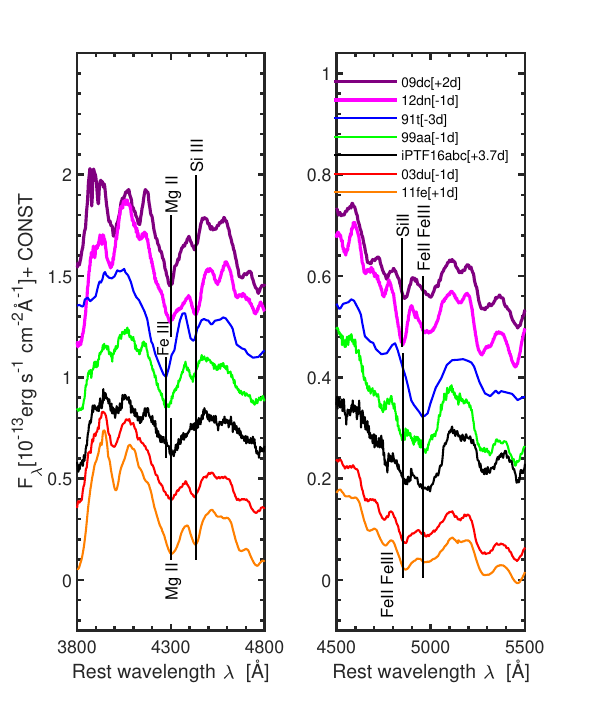}
 \caption{Left: The 4300 \AA\ region shows \FeIII\ absorption characteristic of 91T-like SNe, whereas \MgII\ dominates this feature in normal and 03fg-like SNe. The \SiIII\ contribution remains consistent across all types. Right: The 5000 \AA\ region distinguishes 03fg-like events, where the blue component is driven by \SiII\ $\lambda\lambda$5041, 5056, unlike the \FeII/\FeIII\ blend seen in normal and 91T-like objects. This feature serves as a temperature indicator: the two components are well-separated in 03fg-likes but gradually blend as temperature increases, fusing into a single absorption in the hottest SN\,1991T.
 }
 \label{comp_Fe_lines}
\end{figure}

The feature near 5000 \AA\ remains split into two distinct components, with the separation becoming more pronounced. The redder wing, initially influenced by \SII\ \lam\ 5212, in the pre-maximum epoch, is now entirely dominated by \FeIII\ \llamb\ 5156, 5127, while the blue wing continues to be attributed to \SiII\ \llamb\ 5055, 5041. This behavior differs from both normal and 91T-like supernovae, where this feature typically appears as a broad, deep component characterized by a multiplet of absorptions due to \FeII\ and \FeIII\ lines. A comparison of this region with other SNe is presented in Fig. \ref{comp_Fe_lines}.

\begin{figure}
\includegraphics[trim={0 15 10 22},clip,width=0.35\textwidth]{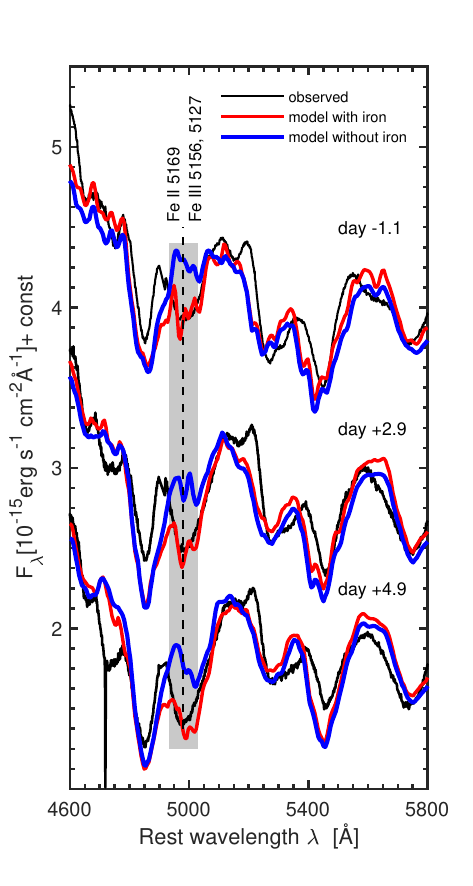}
 \caption {Probing stable iron in the intermediate shells. A stable iron abundance at the 10 per cent level (increasing from 7 per cent to 21 per cent across the shells) is required to reproduce the absorption feature near 5000\AA, which is dominated by a complex of \FeII\ and \FeIII\ lines.}  
\label{probing_iron_middleshells}

\end{figure}

 The stable Fe (\ie\ \Feff) abundance gradually increases from 0.07 at $v > 11,600$\,\kms\ to 0.21 at $v > 9,600$ \kms. \cha{ This is necessary to reproduce accurately the absorption feature near 5000 \AA, as shown in Fig. \ref{probing_iron_middleshells}}. The remaining iron originates from \Nifs\ decay. 
 The fraction of \Nifs\ required increases to 0.45 in the shell between 9,500 and 10,200 \kms.

\textit{Calcium:} The feature near 3800 \AA, typically attributed to high-velocity (HV) \CaII\ H\&K in normal SNe\,Ia, deepens at this epoch. However, it remains a blend of two distinct components: a blue component from \SiII\ \lam\ 3856, 3863 and a red component from photospheric-velocity (PV) \CaII, maintaining the structure observed in the pre-maximum phase. 
In contrast, normal and 91T-like events exhibit deep HV \CaII\ H\&K at the same epoch. The \CaII\ NIR becomes stronger and splits into two components, accurately reproduced in our synthetic spectra. Notably, no calcium was required in these layers to achieve these fits.

\textit{Silicon, Sulphur, Magnesium:} 
 Both \SiII\ \lam\ 6355 and \SiII\ \lam\ 5972 deepen, with their ratio $\mathcal{R}(\text{Si II})$ still being accurately reproduced. The \SiIII\ \lam\ 5739 line, located blueward of \SiII\ \lam\ 5972, weakens progressively until it disappears entirely by day $+$2.9.

The region near 4400 \AA, which previously exhibited two distinct components, now shows a single, deep absorption. On the red side, a blend of \FeII\ and \FeIII\ lines now dominates, replacing the \SiIII\ $\lambda\lambda$4553, 4568 doublet that characterized the earlier spectra. These iron-group transitions, along with the \MgII/\FeIII\ complex on the blue wing, collectively shape this broad and prominent feature.

As in the earlier epochs, the blue component of the 5000 \AA\ feature (centred near 4800\AA) is dominated by a blend of \SiII\ $\lambda\lambda\,$5041, 5055.
This narrow absorption is absent in normal or 91T-like SNe at similar epochs; in those events, the region typically exhibits a broad, deep absorption feature formed by blends of \FeII\ and \FeIII\ lines, as shown in Fig.~\ref{comp_Fe_lines}. This distinctive ``double narrow'' absorption profile is a notable characteristic of SN\,2012dn and other 03fg-like SNe \citep{hicken_2006gz}. 

The $W$-shaped \SII\ feature near 5400 \AA\ and the \SII-dominated 4500--4700 \AA\ region remain well reproduced. At this epoch, a blend of \SII\ \lam\,8859 and \CaII\ \llamb\,8927, 8912 flattens the emission peak redward of the \CaII\ NIR triplet (Fig.~\ref{daymax2}, day $+$2.9), also reproduced by our models. 


\MgII\ continues to contribute to the absorption feature near 4400 \AA, though the line is increasingly influenced by \FeIII\ at this epoch. 
The \MgII\ features around 7500 and 8800 \AA\ remain accurately replicated. Notably, no magnesium is required in these layers, which is consistent with the chemical composition predicted by the W7 model \citep{nomoto1984, iwamoto1999}.

\textit{Carbon, Oxygen:} The \CII\ \lam\ 6580 remains clearly visible near maximum light whereas the \lam\ 7234 feature is less prominent, a behavior that is faithfully reproduced in our synthetic spectra. The carbon mass fraction is sustained at 0.05 down to $v > 9,500$ \Kms\ with oxygen also persisting in these layers. Our derived carbon fraction is an order of magnitude higher than the $X_{\rm C} \approx 0.002$ found in normal SNe\,Ia \citep{Tanaka2011} and falls within the same range as the carbon required for SN\,2009dc \citep{hachinger_2009dc}.

\subsection{The spectra post-maximum light}

\begin{figure*}
 \label{spect6-14}
\captionsetup[subfigure]{labelformat=empty}
	\centering

 \begin{subfigure}{1\textwidth}
		\includegraphics[trim={35 5 52 14},clip,width=0.78\textwidth]{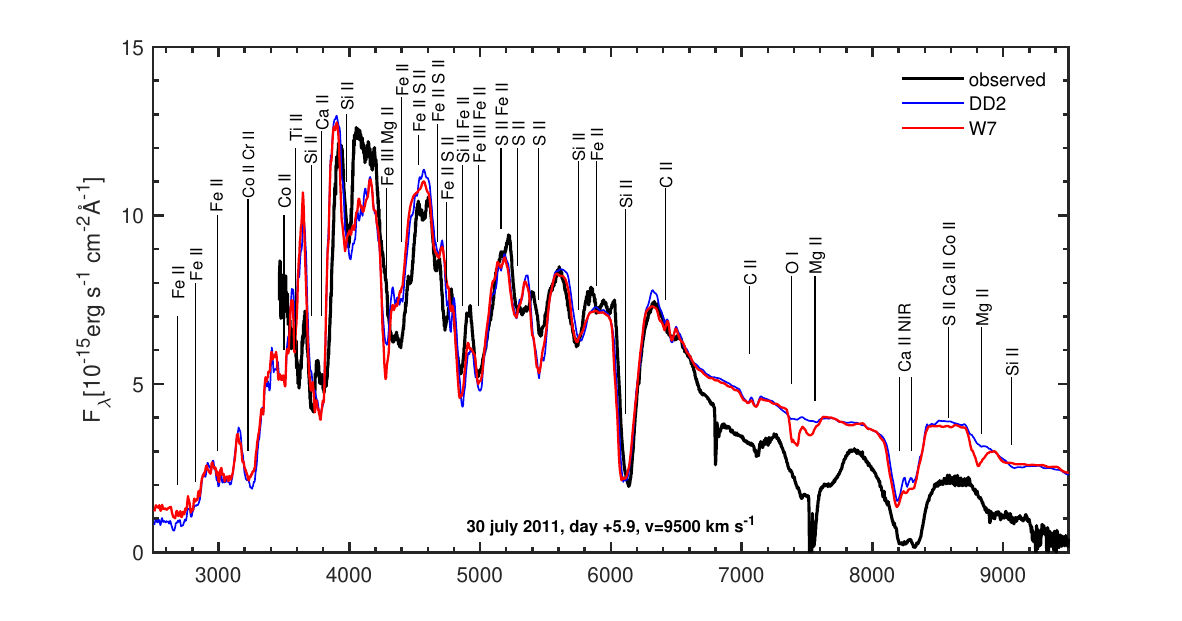}
        \centering
		\caption{ }
	\end{subfigure}
 \vspace{-35pt}

 \begin{subfigure}{1\textwidth}
		\includegraphics[trim={35 1 52 14},clip,width=0.78\textwidth]{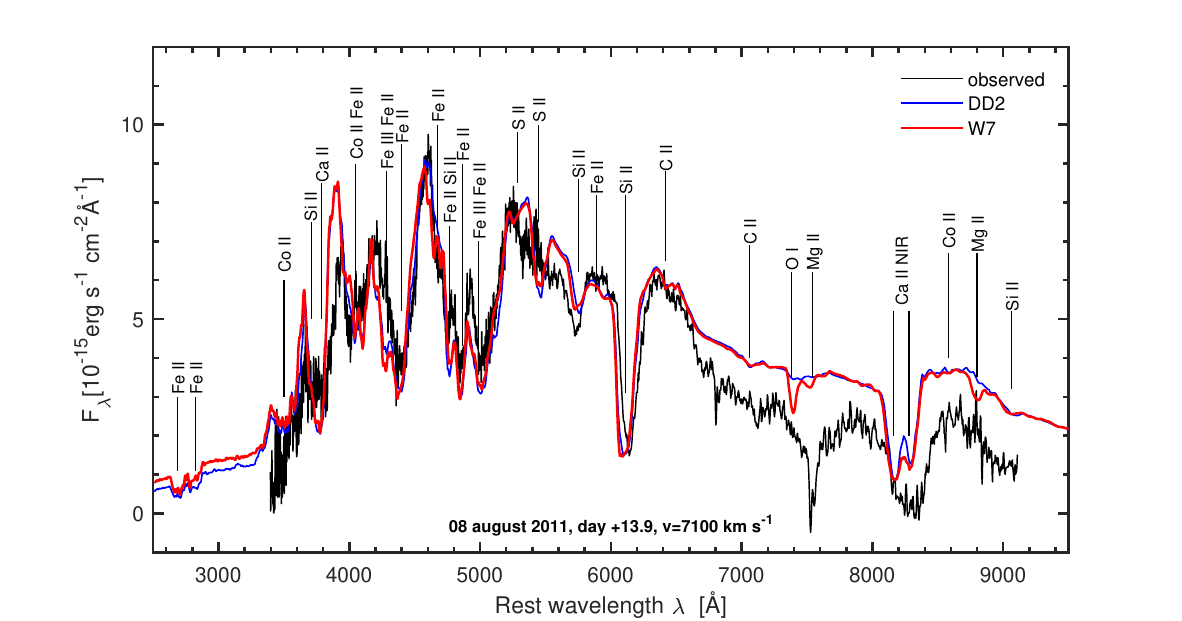}
        \centering
		\caption{ }
	\end{subfigure}

	
\vspace{-20pt}
 \caption{  Spectra one week after maximum light 
 of SN\,2012dn overlaid with the synthetic spectra computed using the W7 and the DD2 density profiles depicted in red and blue, respectively. 
	}
 \label{daypost}
\end{figure*}

In Figure \ref{daypost}, we present synthetic spectra for days $+$5.9 and $+$13.9, overlaid on the observed spectra. The photospheric velocity decreases from 9,500 \kms\  to 7,100 \kms\ . At this epoch, the photosphere has receded into the \Nifs-dominated region, rendering the assumption of a sharp photospheric boundary inadequate. In reality, a significant fraction of the radioactive energy is deposited above the photosphere, a physical effect that is not captured by our model's assumption of a blackbody-emitting surface located at the photosphere. Consequently, this discrepancy results in a synthetic flux excess at wavelengths beyond 6000 \AA.  However, this assumption does not impact the formation of the lines, neither the ionization conditions in the ejecta.

\textit{Fe-group elements:} The feature near 4400 \AA\ becomes increasingly dominated by blended absorptions of \FeII\ and \FeIII, while the influence of \MgII\ and \SiIII\ weakens gradually. 

The feature near 5000 \AA\ remains split into two distinct components: the blue wing still shows a contribution from \SiII, while the red component is now entirely dominated by \FeIII\ and \FeII. 
This structure resembles normal SNe\,Ia \citep[see Fig. 3 of][]{Tanaka2011}, contrasting with 91T-like events where the feature appears broader and lacks two distinct components. In those cases, higher ionization levels mean the feature is not influenced by \SiII, leading to a different spectral morphology \citep{Sasdelli2014, Aouad_99aa}.

By two weeks post-maximum, the absorption feature near 4000 \AA\, initially dominated by \SiII\, becomes increasingly blended with \FeII\ and \CoII\ transitions (see Fig. \ref{daypost}). Similarly, the emission peak on the red wing of the \CaII\ NIR triplet evolves to become entirely dominated by a blend of \CoII\ lines by day $+$14.

The derived \Nifs\ mass fraction in these layers does not exceed 0.45, while the total iron abundance is $\sim$0.27. At these epochs, approximately half of the iron originates from the decay of \Nifs, with the remainder produced by direct explosive nucleosynthesis. 
 
\textit{Calcium:} The feature near 3800 \AA\ shows little evolution, remaining dominated by photospheric \CaII\ in its redder component with a minor \SiII\ contribution on the blue side. Similarly, the \CaII\ NIR triplet maintains a clear double-component profile. While these specific layers require no calcium to fit the current absorption (which originates from higher-velocity material), a small abundance (fractional percentage) is necessary in these deeper shells to account for later nebular-phase emission.

\textit{Silicon, Sulphur, Magnesium:} 
At this phase, the \SiII\ \lam\ 6355 and \lam\ 5972 lines continue to deepen. The model maintains its accurate reproduction of the $\mathcal{R}(\text{Si})$ ratio, confirming that the ionization balance and temperature structure established in earlier epochs remain consistent as.
However, the modeled \SiII\ \lam\ 6355, line velocity could not be fully matched to the observed one, likely due to the development of a blend of emerging \FeII\ lines on the blue wing of the \SiII\ feature, a behavior also observed in SN\,2009dc \citep{hachinger_2009dc}.

The blue component of the feature near 4800 \AA\ continues to be shaped by a blend of \SiII\ $\lambda\lambda$5055, 5041 and \FeII, sustaining the same behavior observed from pre-maximum epochs through this post-maximum phase. 

The absorption near 4000 \AA, which was due to \SiII\ \lam\ 4130 in earlier spectra, is now dominated by \FeII\ and \CoII\ by day $+$14. Similarly, the characteristic $W$-shaped \SII\ feature near 5400 \AA\ remains well-defined at day $+$6; however, it gradually fades over the following week, and is barely discernible by day $+$14. 
Several \FeII\ lines shape the region 4500--4700 \AA\, formerly shaped by blends of \SII\ lines.

As in normal SNe\,Ia, \MgII\ at 4400 \AA\ gradually weakens, being fully replaced by \FeII\ and \FeIII\ by day $+$14. However, \MgII\ remains detectable redward of \OI\ near 7500 \AA\ and in the 8800–9000 \AA\ region. This latter feature, positioned redward of the flat-topped 8500 \AA\ profile, is formed by \MgII\ \llamb\ 9218, 9244.

\textit{Carbon, Oxygen:} Carbon remains necessary ($X_{\rm C} = 0.05$) between $7100 < v < 9100$\,\kms\ to replicate the observed \CII\ \lam\,6580 and \lam\,7234 features. This is supported by the \CII\ \lam\,6580 velocity evolution, which drops from $\sim$12,500\,\kms\, to $\sim$5900\,\kms, confirming that unburned material persists at these lower velocities. Oxygen is similarly required in comparable abundances to fit the \OI\ \lam\,7774 feature.

\begin{figure*} 
\includegraphics[trim={2 2 10 20},clip,width=0.65\textwidth]{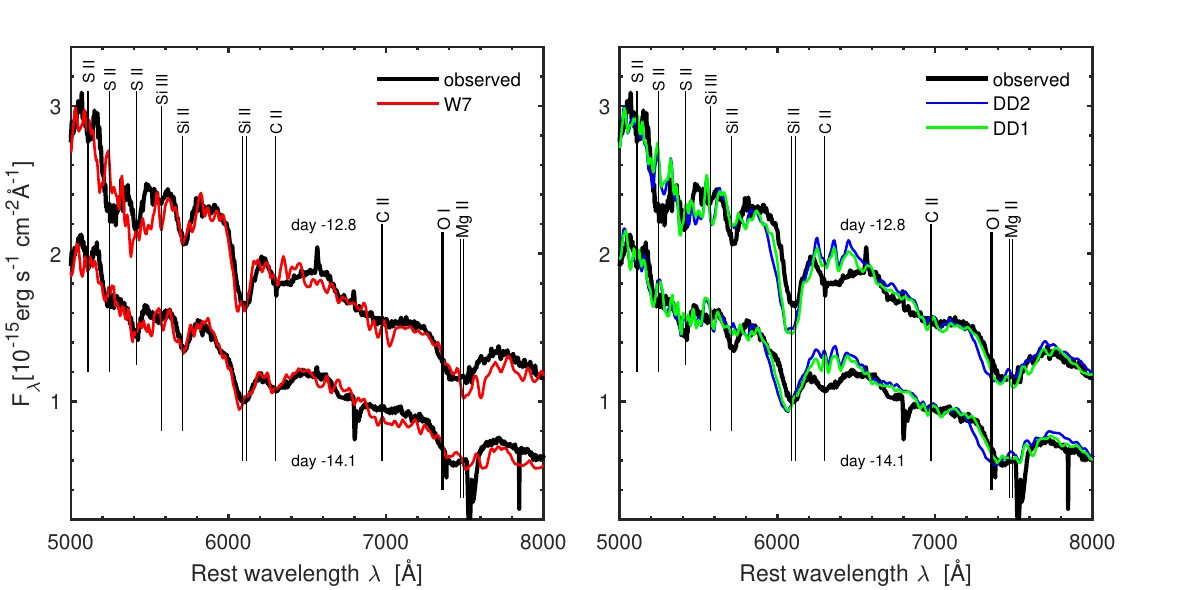}
 \caption{Comparison of observed and synthetic spectra at days $-14$ and $-13$ is presented. The W7 model successfully reproduces the narrow \SiII\ feature near 6000 \AA, the \MgII\ \OI\ blend near 7400 \AA, and the Si II feature near 5800 \AA, which the DD models fail to capture.}
 
   
\label{w7_DD-densitybump}
\end{figure*}

\begin{figure*} 

\includegraphics[trim={0 0 2 2},clip,width=0.7\textwidth]{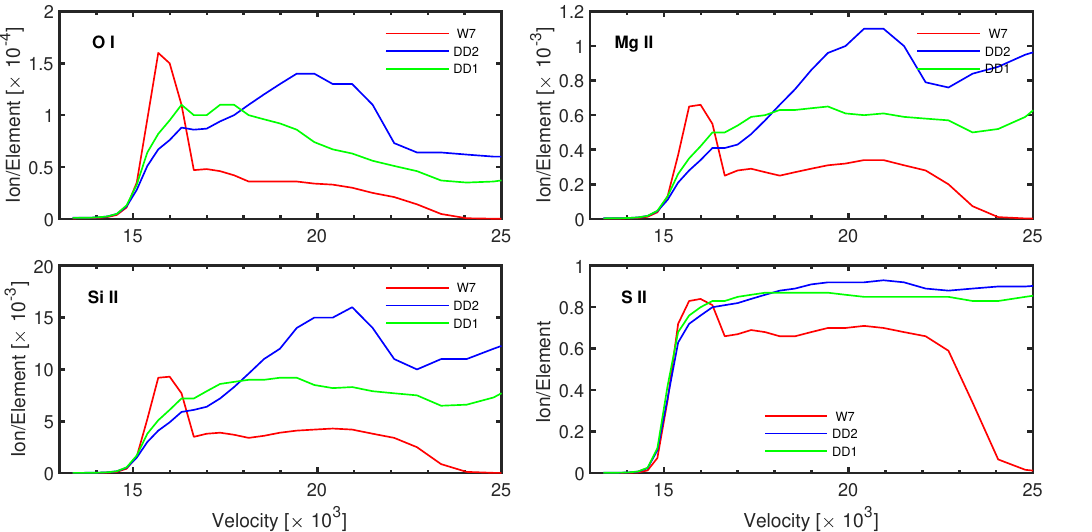}
 \caption{Ionization conditions in the ejecta at day $-13$. In the W7 model, the density bump enhances recombination and lowers the overall ionization level, confining singly ionized Mg, Si, and S, as well as neutral O, within a narrow velocity region around $\sim$16,000 \kms. In contrast, the DD1 and DD2 profiles distribute these ions over a broader velocity range, causing line formation to occur across a more extended region of the ejecta and producing correspondingly bluer line wings (see Fig.~\ref{w7_DD-densitybump}).}
\label{ionization}
\end{figure*}

\subsection{Comparison Between W7 and DD profiles: Density structure and its effect on line formation}
Comparing the W7 model with delayed-detonation models (DD1, DD2) reveals differences in IME feature morphology. As shown in Fig. \ref{w7_DD-densitybump}, the DD models produce overly broad and blueshifted absorptions because their extended outer density structures lower the ionization at high velocities, shifting line formation outward.

In contrast, W7 better reproduces these profiles. Its density enhancement at 12,000–16,000 \Kms\ (Fig. \ref{densityprofiles}) promotes recombination, effectively confining these species to a narrower velocity shell. Ionization analysis (Fig. \ref{ionization}) confirms that the ion fractions of \SiII, \MgII, \SII\ and \OI\ peak sharply within this density bump, resulting in the more accurate, narrower absorption troughs observed in the synthetic spectra, mirroring the behavior of the observed spectra with remarkable precision.

\section{THE NEBULAR PHASE}
\label{nebular_phase}

\begin{figure}

\includegraphics[trim={0 0 5 20},clip,width=0.48\textwidth]{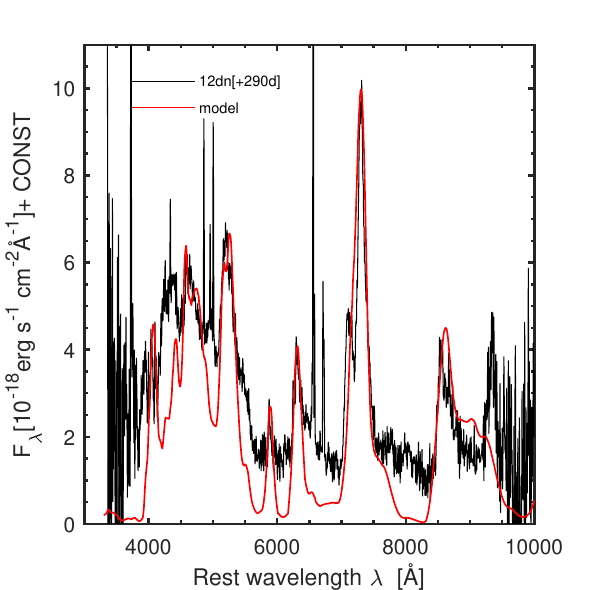}
 \caption{Nebular spectrum at 290 days after B maximum light. The red line is the model.}  
\label{nebular}

\end{figure}

The nebular spectrum of SN\,2012dn is unlike most SNe\,Ia. It is unusual by not showing strong [\FeIII] emission, near 4850\,\AA\ in particular, but it is even more unusual by showing [\OI] \llamb\ 6300, 6363, a feature rarely observed in SNe\,Ia 
This comes together with the unexpectedly low luminosity of the SN at late times, a property that SN\,2012dn shares with one of the ``prototypical'' 03fg-like or ``Super-Chandra'' SNe, SN\,2009dc \citep{taubenberger_2011_2009dc}.

Despite these issues, we attempted to model the spectrum with our nebular code in order to complete the tomographic description of the ejecta. Our code was described, \eg\ in \citet{Mazzali_2020_SN2014J_nebular}. Starting from a density and abundance distribution, it first computes the emission and diffusion of $\gamma$-rays and positrons released by the decay of \Nifs\ and \Cofs. These are assumed to encounter an opacity of 0.027 and 7\,g\,cm$^{-2}$, respectively, and are diffused using a Monte Carlo scheme \citep{Cappellaro1997}. They are assumed to release all of their energy when interacting, \stefan{which} is then transformed into heat, following, \eg\ \citet{Axelrod1980}. This heating is then balanced by cooling via mostly emission lines, which gives rise to the observed spectrum. As many strong lines are in the optical domain, the emission line spectrum is typically strongest in the optical. 
Our model was developed taking into account  two pieces of evidence: the [\OI] emission at low velocity and the sharp dimming of the light curve after $\sim$ day 100. The first element calls for a small mass of oxygen at $v \lsim 4000$\,\kms. Explosion scenarios suggest that low-velocity oxygen may be dragged from a stripped or partially destroyed companion in a double degenerate merging event \citep{pakmoe_2010_WDmergers_equalmass, Pakmor_2012_normalIa_from_merger, kromer_2013_merger} . This material would be additional to any SN ejecta.

The observed dimming appears roughly constant across all photometric bands, consistent with wavelength-independent (grey) extinction. 
We quantify this effect using two distinct methods: first, by comparing the weighted flux attenuation for each individual band relative to the normal SN\,2003du at day 290 (see Fig. \ref{UBVRI_12dn_03du}), and second, by independently evaluating the late-time bolometric light-curve trends (see also Fig.~4 of \citet{taubenberger_2011dn}). Both approaches yield a consistent dimming factor between 4 and 5, the detailed derivation of which is provided in Appendix \ref{grey_dimming}.
\referee{If this grey attenuation is indeed caused by dust, asymmetric nebular emission-line profiles, with the red side appearing weaker than the blue, might be expected. The absence of such pronounced asymmetries in SN\,2012dn, however, does not by itself rule out dust formation, as they are primarily expected for dust distributed throughout the inner ejecta, whereas dust confined to a shell or optically thick clumps may produce little or no line-profile asymmetry \citep{Lucy_dust,Bevan_Barlow_2016}.}

We therefore constructed two models with a \Nifs\ mass four, and five times what is necessary to fit the observed \stefan{nebular} spectrum. \stefan{The resulting spectra were then scaled down by these factors}. Additionally, some mass (0.3 \Msun) was added at low velocities below 7000\,\kms. The
total mass is now 1.66\,\Msun. Oxygen is $\sim$ 20 per cent of the material that was added, \Nifs\ \stefan{is} about 30 percent, the rest being composed of
intermediate-mass elements (IME) which were necessary to preserve the overall
look of the synthetic spectrum. The total \Nifs\ mass in this  model is 0.45--0.49\,\Msun. The result \stefan{using a factor of four} is shown in Fig. \ref{nebular}. \stefan{The model using a factor of five is indistinguishable.}

 
Another SN\,Ia which is known to have shown [\OI] emission at late times is SN\,2010lp \citep{Taubenberger_2013}. This was a low-luminosity, SN\,1991bg-like SN at early times. The nebular spectra could be best reproduced with a two-component model containing $\approx 0.18$\,\Msun\ of \Nifs, consistent with the low peak luminosity \citep{mazzali_2010lp}. The ejecta were consistent with a sub-Chandrasekhar mass. A low-mass progenitor white dwarf would have low central density and would not synthesize neutron-rich iron-group elements. Oxygen was located at low velocities, and an oxygen mass of 0.04\,\Msun\ was sufficient to match the observed [\OI] 6300, 6363\,\AA\ emission line. This may be surviving oxygen from a double-degenerate WD merger episode, although detailed models of a massive merger suggest a larger mass \citep{kromer_2013_merger}. 
Interestingly, SN\,2010lp showed distinctly multiple emission peaks, supporting a DD origin for the event. Such multiple profiles are not observed in SN\,2012dn, but the amount of oxygen is consistent with the high-mass DD model of \citet{kromer_2013_merger}.
Other instances of [\OI] emission include ASASSN-20jq, attributed to an off-center delayed detonation in a $M_{\mathrm{Ch}}$ WD \citep{BOSE_2025_ASASSN-20jq_o_emission_underlum, hoelfich_2021_ASASSN-20jq}, iPTF14atg \citep{KROMER_2016_iPTF14atg_O_emission}, and within the 03fg-like sub-group, SN\,2022pul \citep{Siebert_2024} and SN\,2021zny \citep{Dimitriadis_2023_2021zny_OI_emission_DDmerger}.


\section{ABUNDANCE TOMOGRAPHY}
\label{abundancediscussion}

\begin{figure*}
   
    \begin{subfigure}{\textwidth}
    \centering
       
        \includegraphics [trim={0 2 0 0},clip,width=0.75\textwidth]{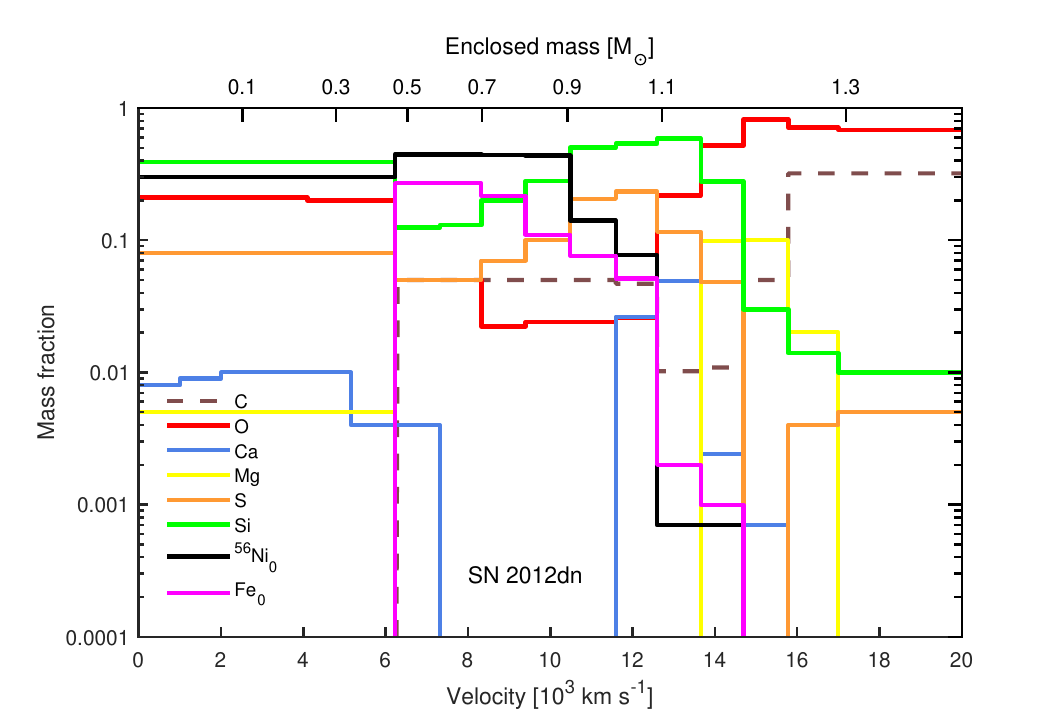}
        
    \end{subfigure}

    \vspace{0.05cm}

    \begin{subfigure}{\textwidth}
    \centering
        
        \includegraphics[trim={0 2 0 0},clip,width=0.75\textwidth]{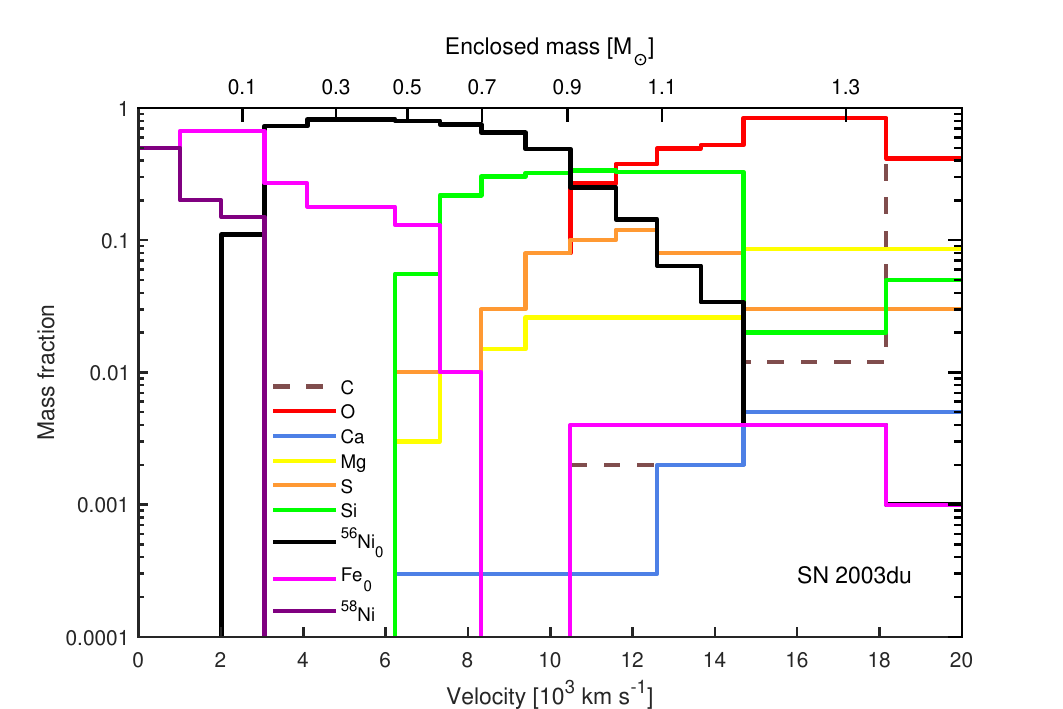}
        
    \end{subfigure}
    
    \caption{Abundance distributions of SN\,2012dn  (upper panel) compared to the normal SN\,2003du. The two events show strikingly different core compositions: while the innermost layers of SN\,2003du are dominated by neutron-rich iron-group elements and are entirely depleted of oxygen, SN\,2012dn features a central oxygen reservoir. In this model, oxygen extends to the core, intermixed with silicon and sulfur, alongside a notably low \Nifs\ abundance.}
    \label{abundance}
\end{figure*}

The yields derived from our modeling are summarized in Table \ref{nucleo_yield} and shown in Figure \ref{abundance}. 
The distribution of abundances is not very different from a normal SN\,Ia but it exhibit\stefan{s} large-scale mixing. Oxygen extends deep into the core, reaching a total mass of $\sim 0.33\,$ \Msun, with approximately $0.1\,$ \Msun\ located below $3000$\,\Kms, a feature essential for reproducing the [\OI] \ \llamb\ 6300, 6364 emission in our nebular spectra. Carbon similarly persists down to $v \approx 6000\text{ km s}^{-1}$ with a mass of $0.03\,M_{\odot}$. 
\chris{Neutron-rich stable Fe-group elements, including \Feff\ and \Nife, are absent from the core}. Stable iron ($0.11\,\Msun$) is additionally absent from the outermost layers, being instead confined to a shell between $6,000 < v < 13{,}000$ \kms. 
While \Nifs\ extends into deeper layers, it remains relatively poor in the core (at $\approx$30 per cent) and peaks within the same shell as the stable iron. In contrast, silicon ($0.54\,M_{\odot}$) and sulfur ($0.14\,M_{\odot}$) span the entire ejecta. The total burned mass is $\sim 1.25\,M_{\odot}$, only about 10 per cent higher than in normal SNe Ia \citep{mazzali2007}, with a total kinetic energy of $\approx 1.4 \times 10^{51}\text{ erg}$. A comparison with SN\,2003du \citep{Tanaka2011} is \stefan{given} in Figure \ref{abundance}.

\begin{table}
\setlength{\tabcolsep}{3pt} 
\centering
\caption{Nucleosynthetic yields and kinetic energies from the modelling compared to the original hydrodynamic models. Results from other SNe are also shown.}
\begin{threeparttable}
\resizebox{\columnwidth}{!}{
\begin{tabular}{llllllll} 
\hline
  & \Nifs & $\mathrm{Fe}$\,$^a$ & $\mathrm{Ni}_{\mathrm{stable}}$ & IME\,$^b$  &O& IME/Fe$^c$ &  $E_\mathrm{k}$ \\
  & M$_\odot$ & M$_\odot$ & M$_\odot$ & M$_\odot$ & M$_\odot$ &  & $10^{51}$\,{\rm ergs} \\
\hline
2012dn   & 0.45 & 0.11  &  0.000  & 0.68  &  0.33 & 1.21 & 1.4 \\
\hline
original W7  & 0.59 & 0.16  &  0.1220  & 0.24  &  0.143 & 0.27 & 1.30 \\
original DD2 & 0.69 & 0.10  &  0.0540  & 0.33  &  0.066 & 0.39 & 1.40 \\
original DD3 & 0.77 & 0.10  &  0.0664 & 0.25  &  0.056 & 0.27 & 1.43 \\
\hline
iPTF16abc$^d$ (DD2) & 0.76 & 0.20  &  0.0060  & 0.19  &  0.21 & 0.20 & 1.32 \\
1991T$^e$ (DD3)     & 0.78 & 0.15  &  0.0006 & 0.18  &  0.29 & 0.19 & 1.24  \\
1999aa$^f$ (DD2)    & 0.65 & 0.29  &  0.0060  & 0.20  &  0.22 & 0.21 & 1.32 \\
2003du$^g$ (W7)     & 0.62 & 0.18  &  0.0240  & 0.26  &  0.23 & 0.32 & 1.25  \\
2002bo$^h$ (W7)     & 0.49 & 0.27  &  0.0001 & 0.28  &  0.11 & 0.37 & 1.24 \\
2004eo$^i$ (W7)     & 0.32 & 0.29  &  0.0005 & 0.43  &  0.30  & 0.70  & 1.10 \\ 
\hline
\end{tabular}
}
\begin{minipage}{\columnwidth}
\begin{tablenotes}[flushleft]
\footnotesize
\item[$^a$] Stable isotopes excluding $^{56}$Fe;
\item[$^b$] $^{28}$Si + $^{32}$S; 
\item[$^c$] \Nifs\ $+$ \Feff\ $+$ \Nife;
\item $^d$\cite{Aouad_iPTF16abc}, $^e$\cite{Sasdelli2014}, $^f$\cite{Aouad_99aa},\\ $^g$\cite{Tanaka2011}, $^h$\cite{Stehle2005}, $^i$\cite{Mazzali2008}.
\end{tablenotes}
\end{minipage}
\end{threeparttable}
\label{nucleo_yield}  
\end{table}


\section{BOLOMETRIC LIGHT CURVE}
\label{bolometriclightcurvesection}
We construct a pseudo-bolometric light curve in the 3,000--10,000 \AA\ range using photometry data from \citet{taubenberger_2011dn} in the $U, B, V, R, I$ bands. Magnitudes are converted to flux using the flux zero points from \citet{FukugitaFLUXZEROPOINTS}, and dereddened with the extinction curve of \citet{cardelli}, assuming a reddening value of 0.09 (see Sect. \ref{data}). The flux for each epoch is calculated using trapezoidal integration between the central wavelengths of each band. Beyond the $U$ and $I$ bands, the flux is extended to 3,000 \AA\ and 10,000 \AA\ using a flat function similar to \citet{Aouad_99aa}. Bolometric luminosities are then derived using a distance modulus of $\mu = 33.26 \pm 0.20$ mag \citep{taubenberger_2011dn}. \chris{However, 03fg-like SNe are known to exhibit enhanced NIR emission relative to the optical. Therefore, the \textit{UBVRI} pseudo-bolometric light curve should be regarded as a lower limit to the true bolometric luminosity \citep{ashall_2021_2003fg_superchandra}}.

\begin{figure}
\includegraphics[trim={25 10 40 35},clip,width=0.5\textwidth]{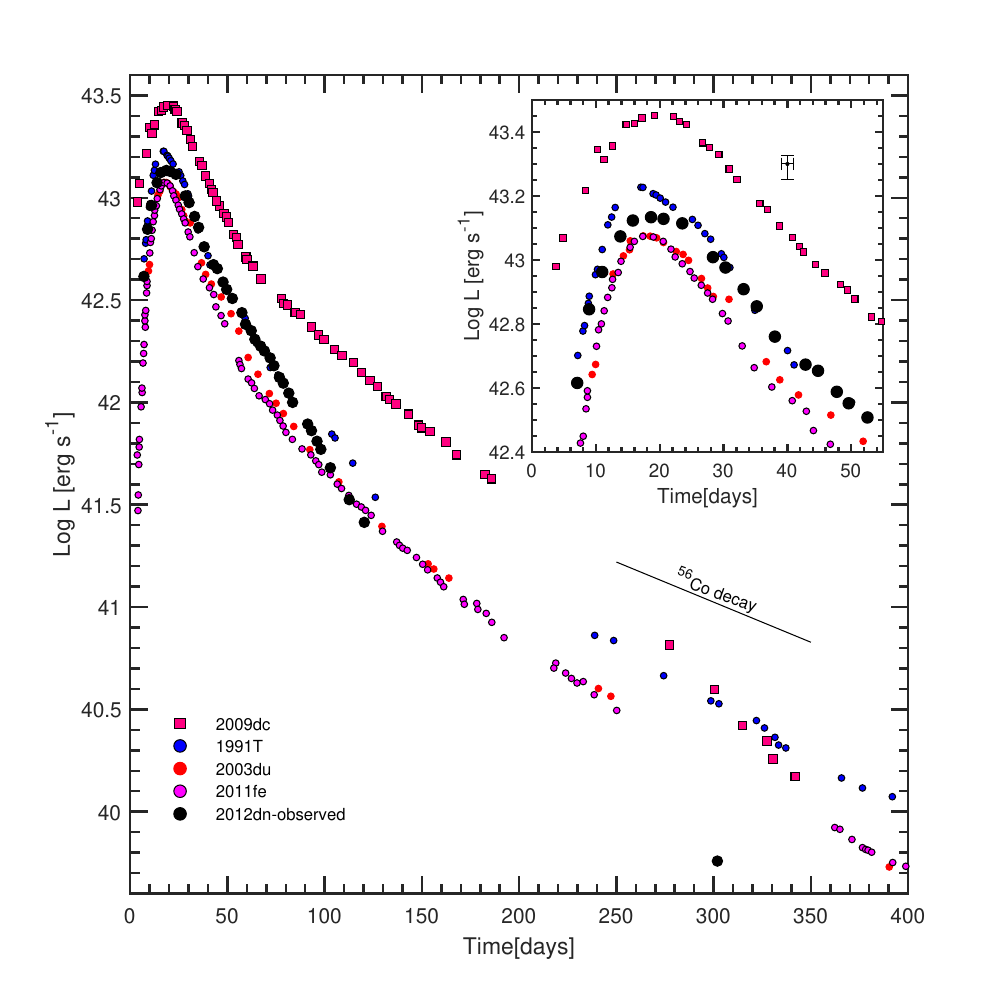}
 \caption{\textit{UBVRI} pseudo-bolometric light curve of SN 2012dn compared with SNe 1991T, 2003du, 2011fe, and 2009dc \citep{Sasdelli2014, stanishev2003du, zhang2016_2011fe, taubenberger_2011_2009dc}. By $\sim$60 days post-maximum, SN 2012dn undergoes a sharp decline, becoming significantly less luminous than the rest of the sample by day 300. SN 2009dc exhibits a similar drop $\sim$100 days later; notably, both events follow a nearly identical, steep decline slope. The vertical error bar reflects the peak luminosity range of SN\,2012dn based on the minimum and maximum distance moduli in the literature.
 }  
\label{Lbol_comparison}
\end{figure}

The supernova reaches a rest-frame peak luminosity of $L_{\mathrm{peak}} = 1.35^{+0.27}_{-0.22} \times 10^{43}$ erg s$^{-1}$ with a rise time of 18.44 days. This peak is \cha{in the same range as}  the transitional SN\,1999aa \citep{Aouad_99aa} and the normal  SN\,2003du. Even assuming the maximum literature reddening of $E(B-V) = 0.18$; \citep{chakdrahari_2012dn}, the resulting peak of $1.78^{+0.27}_{-0.22} \times 10^{43}$ erg s$^{-1}$ remains significantly fainter than extreme 03fg-like events like SN\,2007if and SN\,2009dc, which reach peak luminosities nearly twice that of SN\,2012dn \citep{Taubenberger_2013, ashall_2021_2003fg_superchandra}. By day $+60$, the bolometric decline of SN\,2012dn begins to deviate from that of normal SNe\,Ia, which typically follow a slope slightly steeper than the $^{56}\mathrm{Co}$ decay rate. Notably, while SN\,2009dc exhibits a similar decline, it does not do so until day $+200$.
Comparisons with other SNe\,Ia are provided in Fig.~\ref{Lbol_comparison}, and various literature reports for the peak bolometric luminosity of SN\,2012dn are summarized in Tab.~\ref{bolL}.

\begin{figure}
\includegraphics[trim={25 20 25 25},clip,width=0.49\textwidth]{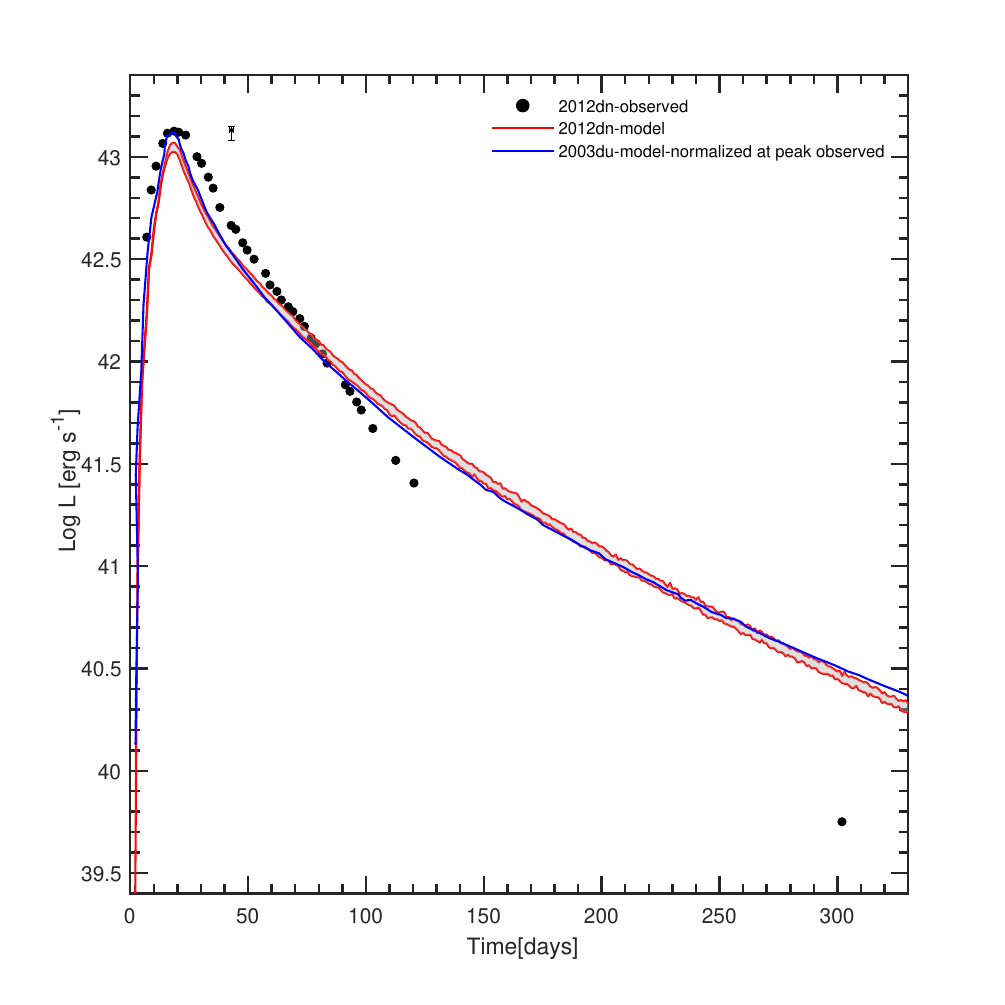}
 \caption{Observed \textit{UBVRI} pseudo-bolometric light curve of SN\,2012dn (black dots) versus synthetic models (red lines) for $M(^{56}\text{Ni}) = 0.45$--$0.49\,M_{\odot}$. The normal SN\,2003du model \citep{Tanaka2011}, scaled to the observed peak (dashed blue), reveals an extinction of $\sim$0.7 dex by day $+$300, highlighting significant late-time luminosity deviation of SN\,2012dn. Despite their shared morphology, the model for SN\,2012dn is broader than that for SN\,2003du, reflecting its larger total ejecta mass.}  
\label{Lbol}
\vspace{-11.5pt}
\end{figure}

Using our abundance results, we construct a synthetic bolometric light curve using a Monte Carlo radiative transfer code \citep{Cappellaro1997, mazzali2001}. The code computes $\gamma$-ray and positron deposition from \Nifs\ decay \citep{Axelrod1980} and follows the diffusion of optical packets through the ejecta. Opacity is dominated by line transitions, parameterized by local Fe-group abundances and temperature \citep{PauldrachNLTElineblocking, mazzali2006Podsiadlowski}. 
As expected, the model fails to reach the observed peak luminosity (Fig. \ref{Lbol}). This discrepancy suggests that an additional energy source may be at play.
For comparison, we plot the light curve model of SN\,2003du \citep{stanishev2003du} normalized to our model peak. The two light curves exhibit nearly identical behavior, with the primary difference being the broader width of SN\,2012dn. This increased width, a direct consequence of the larger ejecta mass ($1.66\,$ \Msun), provides a better fit to the observed evolution than a standard Chandrasekhar-mass profile.
\begin{table}
\setlength{\tabcolsep}{2pt}
\caption{Bolometric luminosity of SN\,2012dn in the $U,B,V,R,I$.}
\label{tab2} 


\scalebox{0.95}{
\hskip-0.5cm\begin{tabular}{llcccccc}

\hline
   Author   &  $\log L_{\mathrm{peak}}$ & Distance modulus & $E (B-V)$  \\
          &    $\mathrm{erg\,s^{-1}}$      &            & \\
  \hline
 
\citet{taubenberger_2011dn} & 43.075 & 33.26  & 0.10  \\
\citet{chakdrahari_2012dn}  & 43.17 & 33.15  & 0.18  \\
This paper & 43.12 & 33.26  & 0.09  \\

\hline
\end{tabular}}
\label{bolL}
\small

\end{table}

\section{VIABLE PROGENITOR-EXPLOSION CHANNELS}
\label{discussion}

To power the peak of SN\,2012dn via radioactive decay alone, $\sim 0.65\,M_{\odot}$ of \Nifs\ is required, based on its peak luminosity similarity to SN\,1999aa and SN\,2003du \citep{Aouad_99aa, Tanaka2011}. Our nebular model initially yields $0.45-0.49$ \Msun\ of \Nifs, a range dictated by the uncertainty of the late-time grey-extinction factor.  Allowing for a modest additional $\sim 0.05$ \Msun\ to account for residual continuum discrepancies between the observed and synthetic spectra (Fig.~\ref{nebular}) raises the total to $\approx 0.54$ \Msun, still below the value required to explain the peak luminosity. 
Even when adopting the upper limit of the inferred \Nifs\ mass range, a deficit persists relative to that required to power the peak. This suggests that even within the most flexible range of our abundance tomography, an additional energy source is required to bridge the remaining gap. 
This shortfall is likely reconciled by a supplemental energy contribution from ejecta interaction with a carbon- and oxygen-rich CSM. Such a scenario is frequently invoked for 03fg-like events to explain their enhanced luminosity, persistent carbon features, and low ionization states \citep{hachinger_2009dc, Noebauer_2016__denseC_O_interacion_superchandra, ashall_2021_2003fg_superchandra}.

\begin{figure*}
 \label{spect6-14}
\captionsetup[subfigure]{labelformat=empty}
	\centering

 \begin{subfigure}{1\textwidth}
		\includegraphics[trim={35 5 52 14},clip,width=0.69\textwidth]{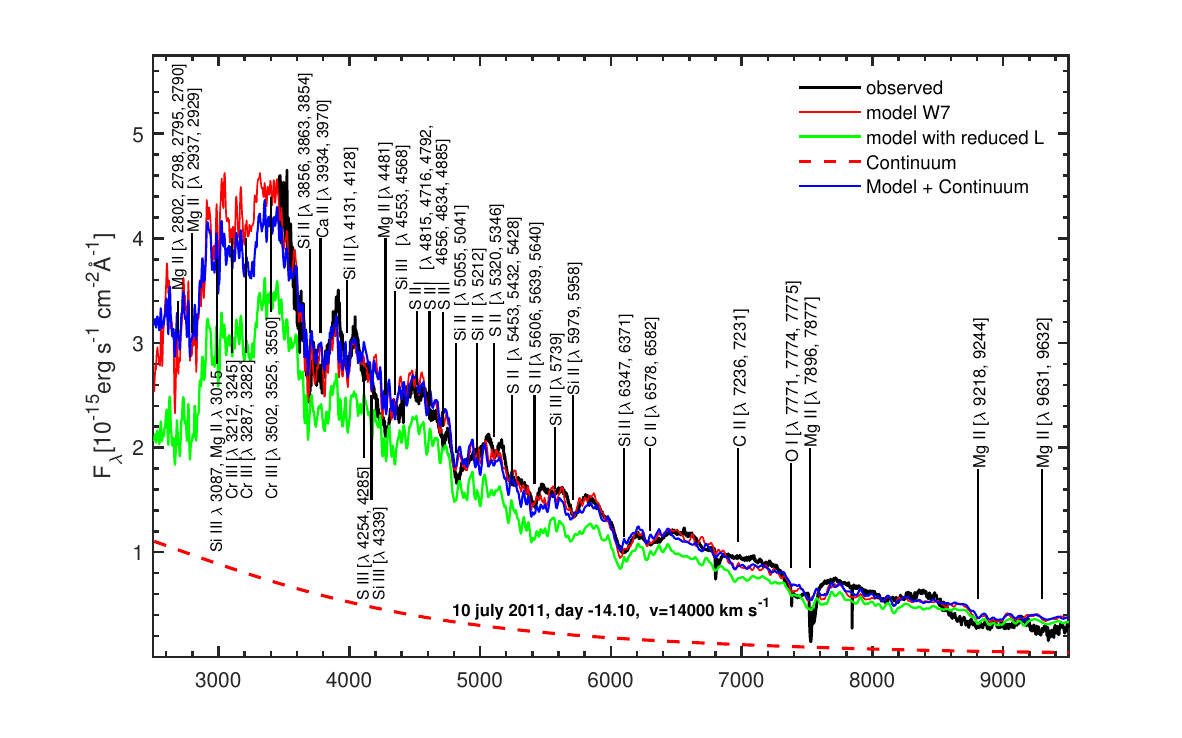}
        \centering
		\caption{ }
	\end{subfigure}
 \vspace{-35pt}

 \begin{subfigure}{1\textwidth}
		\includegraphics[trim={35 1 52 14},clip,width=0.69\textwidth]{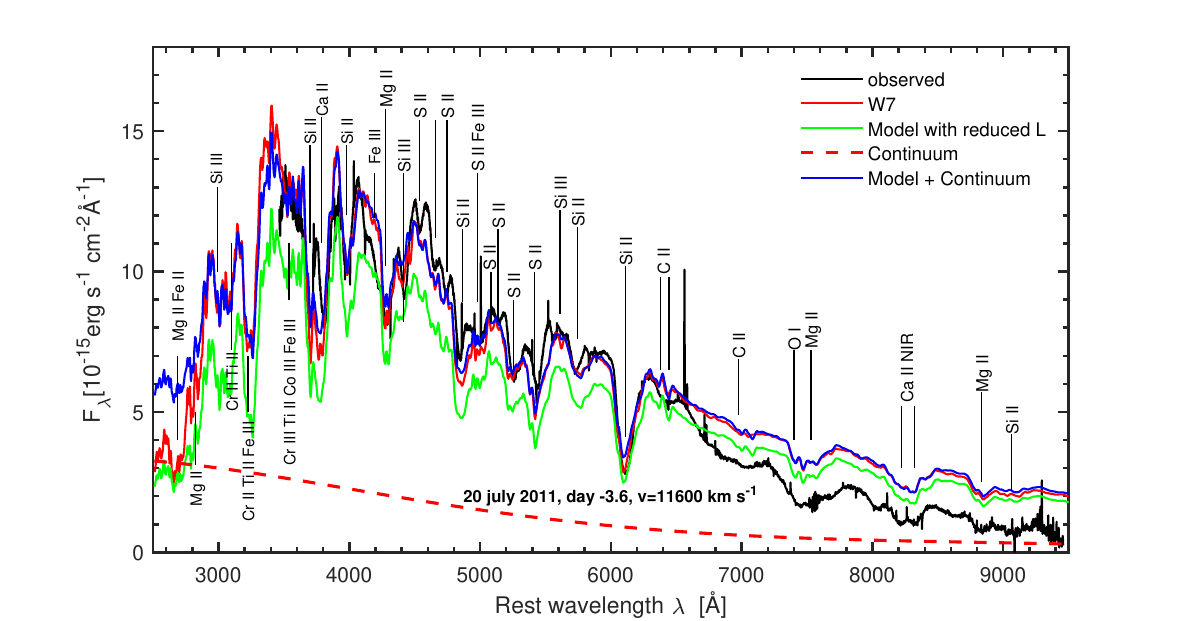}
        \centering
		\caption{ }
	\end{subfigure}

	
\vspace{-20pt}
 \caption{Comparison of observed and modeled spectra incorporating a blackbody (BB) component at two epochs (early and near-maximum light). The observed spectra are shown in black. The initial models are represented by red solid lines, while the models with reduced radioactive luminosity are shown in green. In these final models, the input luminosity was reduced by 25 per cent for the early spectrum and 20 per cent for the near-maximum spectrum. The red dashed lines represent the additional BB component required to account for the total luminosity, with temperatures of $14,000$ K and $11,000$ K for the early and near-maximum epochs, respectively. The blue solid lines represent the total synthetic flux (the sum of the green model and the BB component).}
 \label{spectra_with_BB}
\end{figure*}

Following the methodology of \citet{hachinger_2009dc}, we simulated this effect by reducing the radioactive luminosity in our models and compensating for the flux deficit with a blackbody component. Our analysis at two distinct phases reveals that interaction can account for up to 25 per cent of the total flux at the early epoch, decreasing to  20 per cent near maximum light. Importantly, the inclusion of this component maintains the high quality of the spectral fits (see Fig. \ref{spectra_with_BB}). This requirement is notably less extreme than the 50 per cent early-time contribution necessitated by SN\,2009dc \citep{hachinger_2009dc}. Nevertheless, such a contribution comfortably bridges the gap between our derived \Nifs\ mass  and the observed peak luminosity. These findings suggest a shared physical origin for 03fg-like events, where CSM interaction provides a fundamental, yet highly variable, power source across the class regardless of the specific peak magnitude.

Such an interaction is a \stefan{plausible} consequence of a double-degenerate (DD) merger, a scenario that simultaneously accounts for both the central oxygen core and the circumstellar environment required for SN\,2012dn.
\takashi{Similar observational characteristics have recently been reported for SNe\,2021zny \citep{Dimitriadis_2023_2021zny_OI_emission_DDmerger} and 2022pul \citep{Siebert_2024}, including an early-time flux excess consistent with ejecta–CSM interaction, persistent carbon features, low late-time ionization, reduced stable Fe-group elements abundances, and nebular [\OI] emission, further supporting a double white dwarf merger origin (see also \citealt{Dutta_2022_SN2011aa_DDmerger,Dimitriadis_2022_2020esm_superchandra, Liu_2026_superchandra_2024igg_TARDIS}).} 
\referee{Interestingly, JWST observations of SN\,2022pul provide direct evidence for dust emission \citep{Siebert_2024}, demonstrating that dust formation can occur in 03fg-like Type Ia supernovae. However, despite being only $\sim0.3$ mag fainter than SN\,2012dn in peak absolute $B$-band magnitude, SN\,2022pul remained substantially brighter at late epochs. The substantially larger late-time luminosity deficit observed in SN\,2012dn therefore demonstrates further that the strength of this effect varies significantly among members of the subclass.}


Remarkably, our results resemble the 3D hydrodynamic simulations of a $1.1 + 0.9\,$ \Msun\ CO–CO white dwarf merger by \citet{Pakmor_2012_normalIa_from_merger} to a high degree (see their Fig.~2). Despite the 1D nature of our reconstruction compared to their 3D asymmetric models, the general abundance trends are remarkably consistent: oxygen dominates the core and is surrounded by a shell of stable iron. Silicon extends from the outermost layers deep into the center, \Nifs\ abundance remains poor ($\approx$ 30 per cent) in the innermost layers while carbon persists to intermediate shells. In their simulations, the primary white dwarf’s ashes expand and sweep around the secondary during the burning phase. Because this expansion occurs before the secondary completes its own nucleosynthesis, the secondary’s unburned material eventually dominates the inner ejecta. 
Although the lower-mass merger model of $0.9+0.76 \,$ \Msun\ of \citet{kromer_2013_merger}, predicts nebular oxygen line, it synthesizes a \Nifs\ mass far too low to account for the peak luminosity of SN\,2012dn. 


\chris{However, a discrepancy arises in the innermost composition: our modelling suggests the absence of neutron-rich isotopes such as \Nife\ and \Feff\ in the core, whereas the merger models discussed above, predict measurable \Nife\ under solar-metallicity conditions, as also seen in the equal-mass $0.9+0.9,\Msun$ model of \citealt{pakmoe_2010_WDmergers_equalmass}. However, low neutronization may reflect not only lower densities (and thus progenitor mass) but also low metallicity \citep{Arnett1982}. The absence of neutron-rich ions in the core of SN\,2012dn therefore suggests that both effects act to suppress their synthesis, consistent with our inferred low-metallicity progenitor.}

Constraining the white dwarf mass ratio therefore requires hydrodynamical simulations \chris{and remains further complicated by the possible degeneracy with progenitor metallicity;} nevertheless the CO–CO merger framework remains the most consistent explanation for the multi-faceted properties of SN\,2012dn.

\section{CONCLUSIONS}
\label{conclusions}

We have presented \stefan{our} abundance tomography analysis of the 03fg-like SN 2012dn. This event is highly peculiar, exhibiting a peak luminosity comparable to normal SNe Ia while retaining 03fg-like spectroscopic hallmarks, such as narrow IME features, weak \FeIII\ lines, and persistent carbon. Notably, in the nebular phase,  SN\,2012dn does not show strong [\FeIII] emission, near 4850\,\AA\ and displays  [\ion{O}{i}] emission, rare among type Ia SNe and detected for the first time in this subclass. Furthermore, SN\,2012dn exhibits an unusual wavelength-independent dimming starting $\sim$60 days post-maximum, likely due to grey or clumpy dust formation. Our results can be summarized as follows:

\begin{itemize}
\item The photospheric phase is remarkably well-reproduced by a standard W7 Chandrasekhar-mass density profile. The W7 model outperforms delayed-detonation models (e.g., DD1, DD2) by accurately replicating the narrow absorption troughs. The density enhancement in W7 at $12,000$--$16,000$ \kms\ confines singly ionized species within a narrow velocity shell, preventing the overly broad, blueshifted features produced by DD models.

\item Modeling the nebular phase is inconsistent with a  Chandrasekhar-mass model. Reproducing the [\OI] emission necessitates a low-velocity mass component that brings the total inferred ejecta mass to $1.66\,M_{\odot}$. This total includes $0.33\,M_{\odot}$ of oxygen (of which 0.1\,\Msun\ is concentrated in the core). This oxygen-rich core is largely devoid of stable iron, which instead resides in a surrounding shell. The total \Nifs\ mass, estimated in the range ($0.45$--$0.54\,$\Msun) is modulated by the bolometric factor required to account for the exceptionally low nebular-phase flux caused by grey or clumpy dust extinction. 
Carbon persists down to $v \approx 6000$\,km\,s$^{-1}$, while silicon and sulfur extend from the interior to the surface. 

\item The derived \Nifs\ yield is insufficient alone to power the peak luminosity. Regardless of the specific mass within our derived range, an additional energy source is required to bridge the remaining deficit. This shortfall in radioactive heating is reconciled by supplemental energy from ejecta-CSM interaction, which we model as a thermal blackbody component. By incorporating this component -- accounting for up to 20 per cent of the maximum-light luminosity -- we achieve a high-quality fit that compensates for the nickel mass discrepancy.


\item The derived abundance distribution resembles the outcomes of hydrodynamical models of double-degenerate (DD) mergers. In this scenario, unburned material from the secondary WD is dragged into the center, creating an innermost volume dominated by unburned fuel rather than products of complete nucleosynthesis.

\item The absence of stable iron in the outermost layers of the ejecta constrains the progenitor to sub-solar metallicity. This provides a physical explanation for the unusually blue pre-maximum colors of SN\,2012dn and supports the consensus that 03fg-like events originate in metal-poor environments. \chris{This may imply that these events may be  more common at higher redshift. Given their significant Hubble residuals, they remain a potential source of contamination in future high-redshift cosmological supernova surveys.}

\item This study highlights the necessity of late-time IR observations to recover the full bolometric budget in 03fg-like transients.

\item Hydrodynamic simulations of double-degenerate (DD) mergers involving varying white dwarf mass ratios are required to fully explain the peculiarities observed in SN\,2012dn. Currently, the DD merger remains the most cohesive framework for this event; this progenitor channel not only provides the necessary structure for an oxygen-rich, iron-poor core, but also establishes the environments required for dust formation and ejecta-CSM interaction. Ultimately, the variability in both \Nifs\ production and strength of the supplemental CSM interaction may offer a consistent explanation for the wide range of peak luminosities observed across this class of objects.
\end{itemize}

\section*{ACKNOWLEDGMENTS}
The authors express their gratitude to Stefan Taubenberger for sharing the nebular spectra in a readable format, to Masaomi Tanaka for valuable discussions, and to the anonymous referee for their meticulous review and constructive suggestions, which significantly improved the manuscript.

\section*{Data availability}
The spectroscopic data underlying this article are available at the Weizmann Interactive Supernova Data Repository (WISeREP) \citep{wiserep}.

\bibliographystyle{mnras}
\bibliography{references}

\section{APPENDIX}
\appendix
\section{Estimating of the Bolometric Dimming Factor}
\label{grey_dimming}

The bolometric extinction was estimated by comparing the observed \stefan{\textit{UBVRI} }light curves of SN\,2012dn to those of the normal SN\,Ia 2003du. After normalizing both datasets at peak brightness (See Fig.\ref{UBVRI_12dn_03du}), the systematic offset between the two was interpreted as additional extinction. To determine the total bolometric effect, the extinction in each individual band was first converted into its corresponding transmitted flux fraction. These transmitted fractions were then weighted by the spectral energy distribution (SED) of the \stefan{\textit{UBVRI}} bands, representing the fractional contribution of each photometric band to the total bolometric flux; the specific values adopted for this calculation are summarized in Table~\ref{tab:bol_extinction}. Summing the weighted transmitted fluxes yields an observed bolometric flux of approximately $23$ per cent of the intrinsic value, corresponding to a dimming factor of $4.5 \pm 0.5$, given the inherent uncertainties in the photometric data.
As an independent validation, the nebular-phase bolometric luminosity of SN\,2012dn was compared with that of the normal Type Ia SN\,2003du (normalized at peak). This comparison utilized the light curve from \citet{taubenberger_2011dn} (see their Fig.~4), which suggests an attenuation factor of approximately $0.6$ dex, as well as the pseudo-bolometric light curve computed in this work (Fig. \ref{Lbol}), which yields approximately $0.7$ dex. Both independent approaches provide a consistent attenuation factor range of $4\text{--}5$. %

\begin{table}
\centering
\setlength{\tabcolsep}{4pt}
\caption{Band-dependent extinction values, corresponding flux transmission factors, SED fractions, and weighted transmitted flux contributions used to estimate the bolometric attenuation.}
\label{tab:bol_extinction}
\begin{tabular}{lcccccc}
\hline
 & U & B & V & R & I & Sum \\
\hline
Extinction (mag) & 0.78 & 1.68 & 2.10 & 1.07 & 1.60 &  \\
Transmitted fraction $_a$ & 0.49 & 0.21 & 0.14 & 0.37 & 0.23 &  \\
Bolometric weight $_b$ & 0.02 & 0.45 & 0.24 & 0.15 & 0.14 & \textbf{1.00} \\
Weighted transmission $_{(a \times b)}$ & 0.01 & 0.10 & 0.03 & 0.06 & 0.03 & \textbf{0.23} \\
\hline
\end{tabular}
\end{table}

\bsp	
\label{lastpage}
\end{document}